\documentclass[twocolumn]{aastex63}
\usepackage{multirow}
\usepackage{color}
\usepackage{lineno}

\usepackage{capt-of}
\usepackage{rotating}
\usepackage{longtable}
\usepackage{graphicx}
\usepackage{float}
\usepackage{amsmath,amssymb}
\usepackage{gensymb}
\usepackage{xcolor}
\usepackage{natbib}
\usepackage[hang,flushmargin]{footmisc}
\usepackage{chngcntr}
\usepackage{hyperref}
\usepackage{enumitem}
\setlist{parsep=0pt,listparindent=\parindent}
\usepackage{xtab}
\usepackage{tabularx} % works for 1st table only
\usepackage{xltabular} % works for 2nd table only
\usepackage{soul}
\usepackage{array}

\usepackage{mwe}
\usepackage{afterpage}
\usepackage{rotating}
\usepackage{afterpage}

\usepackage{colortbl}
\usepackage{lipsum}
    {\global\pdfpageattr\expandafter{\the\pdfpageattr/Rotate 90}}%
    {\clearpage\pagebreak[4]\global\pdfpageattr\expandafter{\the\pdfpageattr/Rotate 0}}%

\usepackage{pifont}

\newcommand{\NgridMatsumoto}{3013}

\definecolor{LightCyan}{rgb}{0.88,1,1}

\newcommand{\CfA}{Center for Astrophysics $|$ Harvard \& Smithsonian, Cambridge, MA 02138, USA}

\newcommand{\IAIFI}{The NSF AI Institute for Artificial Intelligence and Fundamental Interactions}

\newcommand{\MIT}{Department of Physics, Massachusetts Institute of Technology, Cambridge, MA 02139, USA}

\begin{document}

\title{Finding the Fuse: Prospects for the Detection and Characterization of Hydrogen-Rich Core-Collapse Supernova Precursor Emission with the LSST}

%\suppressAffiliations

\author[0000-0003-4906-8447]{A.~Gagliano}
\affiliation{\IAIFI}
\affiliation{\CfA}
\affiliation{\MIT}

\author[0000-0002-9392-9681]{E.~Berger} 
\affiliation{\IAIFI}
\affiliation{\CfA}

\author[0000-0002-5814-4061]{V.~A.~Villar} 
\affiliation{\IAIFI}
\affiliation{\CfA}

\author[0000-0002-1125-9187]{D.~Hiramatsu} 
\affiliation{\IAIFI}
\affiliation{\CfA}

\author{R.~Kessler} 
\affiliation{Kavli Institute for Cosmological Physics, University of Chicago, Chicago, IL 60637, USA}
\affiliation{Department of Astronomy and Astrophysics, University of Chicago,  Chicago, IL 60637, USA}

%\author{Tatsuya Matsumoto} 
\author[0000-0002-9350-6793]{T.~Matsumoto}
\affiliation{Department of Astronomy, Kyoto University, Kitashirakawa-Oiwake-cho, Sakyo-ku, Kyoto, 606-8502, Japan}
\affiliation{Hakubi Center, Kyoto University, Yoshida-honmachi, Sakyo-ku, Kyoto, 606-8501, Japan}

\author[0000-0001-8949-5131]{A.~Gilkis}
\affiliation{Institute of Astronomy, University of Cambridge, Madingley Road, Cambridge CB3 0HA, UK}

\author[0000-0003-1009-5691]{E.~Laplace}
\affiliation{Heidelberger Institut f{\"u}r Theoretische Studien, Schloss-Wolfsbrunnenweg 35, D-69118 Heidelberg, Germany}

%\author{Other Folks}

\submitjournal{ApJ}

\correspondingauthor{A.~Gagliano}
\email{alexander.gagliano@cfa.harvard.edu}

%%%%%%%%%%

\begin{abstract}

Enhanced emission in the months to years preceding explosion has been detected for several core-collapse supernovae (SNe). Though the physical mechanisms driving the emission remain hotly debated, the light curves of detected events show long-lived ($\geq$50 days), plateau-like behavior, suggesting hydrogen recombination may significantly contribute to the total energy budget. The Vera C. Rubin Observatory's Legacy Survey of Space and Time (LSST) will provide a decade-long photometric baseline to search for this emission, both in binned pre-explosion observations after an SN is detected and in single-visit observations prior to the SN explosion. In anticipation of these searches, we simulate a range of eruptive precursor models to core-collapse SNe and forecast the discovery rates of these phenomena in LSST data. We find a detection rate of $\sim 40-130$ yr$^{-1}$ for SN~IIP/IIL precursors and $\sim 110$ yr$^{-1}$ for SN~IIn precursors in single-epoch photometry. Considering the first three years of observations with the effects of rolling and observing triplets included, this number grows to a total of 150-400 in binned photometry, with the highest number recovered when binning in 100-day bins for 2020tlf-like precursors and in 20-day bins for other recombination-driven models from the literature. We quantify the impact of using templates contaminated by residual light (from either long-lived or separate precursor emission) on these detection rates, and explore strategies for estimating baseline flux to mitigate these issues. Spectroscopic follow-up of the eruptions preceding core-collapse SNe and detected with LSST will offer important clues to the underlying drivers of terminal-stage mass loss in massive stars.
%a strategy for optimizing our discovery rates by custom templates, and discuss the impact of our results on .

\end{abstract}

\section{Introduction} \label{sec:intro}
Evidence has steadily grown that the formation of circumstellar material (CSM) regularly precedes the death of massive stars as core-collapse supernovae (CCSNe). Persistent narrow emission lines appearing in the spectra of SNe and caused by the photo-ionization of slow-moving ($\sim10-1000$\;km s$^{-1}$; \citealt{1997Filippenko_SNeIIn,2013Taddia_CSPIIn}) CSM by the SN ejecta have long suggested that at least \textit{some} SNe do not explode into pristine environments. The most common sub-type exhibiting these features, SN~IIn \citep[where the `n' stands for `narrow';][]{1990Schlegel_IIn}, is spectroscopically characterized by an early blue continuum and prominent H$\alpha$ emission, and the class exhibits broad photometric diversity dominated by CSM interaction. Balmer emission features lasting weeks to months have also appeared in the spectra of SNe with well-defined photometric plateaus, leading to the designation of the SN~IIn-P sub-class \citep{2013Smith_IInP,2013Mauerhan_IInP,2020Fraser_Interacting}. This long-lived interaction is not limited to hydrogen-rich SNe: narrow spectral features from slow-moving CSM are also observed in hydrogen-poor SNe, including helium-rich SNe~Ibn \citep{2020Sun_SNeIbn,2020Gangopadhyay_Ibnspectroscopy,2022Maeda_SNeIbn,2023Pursiainen_Ibn,2024Wang_SNeIbn} and helium-poor SNe~Icn \citep{2022Pellegrino_Icn,2022GalYam_Icn,2023Davis_2022ann,2023Nagao_2021ckj}.

The detection of prominent emission lines above the SN photosphere requires high CSM densities, and by measuring the velocity of the post-shocked shell from intermediate-width spectral features (or by inferring the shock velocity from light curve models), mass-loss rates in the terminal progenitor system can be estimated \citep[see ][for a review]{2014ASmith_MassLoss}. Sample studies of SNe~IIn have reported progenitor mass-loss rates of 10$^{-2}$ to $>1\;M_{\odot}\;\rm{yr}^{-1}$ \citep{2014ASmith_MassLoss,2017Smith_InteractingSNeIIn}, far exceeding expectations for line-driven stellar winds and ostensibly supporting an eruptive origin with unclear underlying physics. Further complicating the picture, the existence of local CSM surrounding otherwise typical SNe~IIP/IIL has been more recently suggested by the presence of short-lived narrow spectral lines originating from material flash-ionized by the initial pulse of SN photons. Comprehensive searches for these flash-ionization features, paired with the timescales of observed shock breakout, indicate that dense CSM formation is a regular occurrence in CCSNe \citep{2018Forster_SBO, 2024JacobsonGalan_CSMInteractingSNeII}. Stars less massive than $<30\;M_{\odot}$ likely end their lives as red supergiants (RSGs), which are not expected to exhibit eruptive, Luminous Blue Variable (LBV)-like mass-loss. Multiple mechanisms have been proposed to explain this enhanced mass-loss in RSGs, including the damping of convective waves launched by changes in late-stage nuclear burning \citep{2012Quataert_WaveDriven}, pulsation-driven superwinds \citep{1997Heger_Pulsations,2010Yoon_Pulsation} and binary interaction \citep{2024Matsuoka_BinaryInteraction}\footnote{A binary explanation is supported by strong evidence that the majority of massive stars form in binaries with separations low enough to undergo interaction during their lives \citep{1998Vanbeveren_MassiveStars,2012Sana_BinaryInteraction}.}.

%A broad diversity of hypotheses have been proposed to explain this dramatic terminal mass-loss, including binary interaction \citep[e.g.,][]{2024Ercolino_BinaryInteraction} and eruptions from Luminous Blue Variables (the latter hypothesis is bolstered by the direct detection of Luminous Blue Variable (LBV) progenitors in the case of the SNe~IIn SN~2005gl \citep{2009Nature_2005gl} and SN~2009ip \citep{2010Smith_2009ip,2011Foley_2009ip}).

%In spite of its observational support in a few cases, an LBV progenitor for SNe~IIn is difficult to reconcile with standard stellar evolutionary theory, in which the LBV stage follows main sequence burning and cannot immediately precede core-collapse \citep{2000Maeder_LBVs,2011Dwarkadas_LBVProgenitors}. 
%Consequently, it has been suggested that the mass-loss rates inferred from observations of dense CSM are systematically biased by the SN ejecta sweeping up lower-density material \citep{2011Dwarkadas_LBVProgenitors,2024Fuller_BoilOff}.

Complementing the rapid spectroscopic follow-up that has made these discoveries possible, wide-field photometric surveys have both expanded the local discovery volume and increased the typical cadence of observations. The All-Sky Automated Survey for SNe \citep[ASAS-SN;][]{2014Shappee_ASASSN}, the Asteroid Terrestrial Last-Alert System \citep[ATLAS;][]{2018Tonry_ATLAS}, the Zwicky Transient Facility \citep[ZTF;][]{2019Bellm_ZTF,2019Masci_ZTF,2019Graham_ZTF,2020Dekany_ZTFObs}, and the Young SN Experiment \citep[YSE;][]{2021Jones_YSE,2023Aleo_YSEDR1} provide near-synoptic coverage of the sky and temporal sampling of $<$1~day for events serendipitously occurring in the overlap between survey footprints. This wealth of archival imaging has enabled both proactive and retroactive searches for pre-explosion emission from local SNe. This thread began with the detection of emission preceding the Type Ibn SN 2006jc (the precursor light curve is shown as the right panel of Figure~4 in \citealt{2007Nature_Pastorello}; \citealt{2007Foley_Pastorello}).  The detection was followed by precursor emission detected in 2010mc \citep[][light curve shown in Figure~1]{2013Ofek_2010mc}, 2015bh \citep[Figure 2,][]{2017Thone_2015bh}, 2016bdu \citep[erroneously listed as 2016bhu in \citealt{2022Matsumoto_PrecursorModel};][]{pastorello2018supernovae}, LSQ13zm \citep{2016Tartaglia_LSQ13zm}, and 2009ip \citep{2009Berger_2009ip,2011Foley_2009ip,2012ATel_2009ip,2013Mauerhan_2009ip,2014Margutti_2009ip,2013Pastorello_2009ip}, the latter followed three years later by a presumably terminal explosion \citep[although this interpretation is not conclusive;][]{2014Graham_2009ip}. More recently, precursor activity has been detected in the SN~IIn-P 2020pvb \citep{2024EliasRosa_2020pvb}, the normal SN~II 2020tlf \citep{2022Galan_FinalMoments}, the SN~IIn 2021qqp \citep{2024Hiramatsu_21qqp}, and the SN~Ibn 2023fyq \citep{2024Brennan_2023fyq,2024Dong_2023fyq}. These precursors were all detected in individual difference images, but systematic efforts are now underway to search for sub-threshold emission preceding local CCSNe by coadding difference images \citep{2014Ofek_IInPrecursors} or time-averaging photometry in bins leading up to the explosion \citep{2021Strotjohann_MonthsLong}. High-cadence coverage of recent local events have also allowed us to place stringent limits on the luminosity of an optical precursor when an event is not detected either in single-visit difference imaging or through binning (2023ixf, \citealt{2024Ransome_TwilightYears}; 2024ggi, \citealt{202Shrestha_2024ggi}\footnote{Precursor activity is likely in both of these events, with dramatic NIR variability revealed by \textit{Spitzer} in the 2-3 years preceding the former \citep{2023Kilpatrick_2023ixf} and flash-ionization lines indicating the presence of CSM in the latter \citep{2024JacobsonGalan_2024ggi}}).

In this nascent and rapidly-evolving discovery space, several questions loom: is the pre-explosion mass-loss of most terminal systems continuous or eruptive? What physical mechanism drives this emission, and what signatures does it imprint on the timescale and luminosity of the emission relative to the explosion? Is enhanced mass-loss ubiquitous across core-collapse progenitors, and can it be used to identify the final stages of stellar evolution and predict the properties of a subsequent explosion?

Beginning its science operations for the ten-year Legacy Survey of Space and Time (LSST) in late 2025, the Vera C. Rubin Observatory \citep{2019Ivezic_LSST} is expected to discover 1M SNe$\;$yr$^{-1}$, $\sim$50\% of them CCSNe. Its deep ($r\sim$24 and $i\sim$23.4 in a single visit\footnote{\url{https://www.lsst.org/scientists/keynumbers}}) \textit{ugrizY} photometry will enable broad studies of precursor demographics and volumetric rates. These data offer the potential to shift SN science from a \textit{retroactive} discipline (the forensics of stellar death) to a \textit{proactive} one (preparing for and witnessing the most local CCSNe in real-time). A systematic investigation of the anticipated discoveries will better prepare the community for this shift.

In this work, we forecast the discovery rates of eruptive mass-loss episodes from hydrogen-rich CCSNe with the upcoming Rubin Observatory LSST. We separately consider the eruptions preceding SNe~IIn and SNe~IIP/IIL, and consider both theoretical models from previous studies and an observational model constructed from a single archetypal precursor event (that associated with SN~2020tlf). We consider these precursor episodes as standalone transient events without modeling an associated SN, due to the significant uncertainties surrounding their connection to stellar death.

We first explore the potential for characterizing precursor emission that passes LSST's detection trigger in individual differential photometry, which will be possible whether or not a subsequent explosion is observed. We call these `single-visit precursors', and assume the differential photometry is measured from deep templates with zero flux contamination. We also evaluate both the number and properties of precursors recovered by time-averaging multiple epochs of LSST photometry in fixed-width bins, as is common for retroactive searches\footnote{Although we analyze differential photometry in this work, searching for emission in deep coadded images is also common.}. We call these detected events `binned precursors'. This latter technique will allow us to probe pre-explosion variability long before an associated SNe, in contrast with extant detections probing the final months preceding detonation.

Our paper is structured as follows. We describe our precursor models in \textsection\ref{subsec:precursormodel} and the observational model used to generate synthetic LSST data in \textsection\ref{subsec:snana}. In \ref{subsec:singlevisit}, we present our anticipated annual discovery rates in single-visit photometry. We expand this analysis to the first three years of LSST by binning differential photometry in \textsection\ref{subsec:binnedvisits}, and consider strategies to mitigate the impact of flux contamination in our reference photometry \textcolor{black}{in \textsection\ref{subsub:bin_strategies}}. We then discuss the prospects for a local-volume survey for eruptive precursors and their host galaxies in \textsection\ref{subsec:surveyStrategy}, and quantify the impact of our recovery rates on line-of-sight extinction in the case of a dust-enshrouded progenitor in \textsection\ref{subsec:dust}. We summarize our results and conclude in \textsection{\ref{sec:conclusions}}. In all simulations, we adopt a cosmology of $\Omega_M = 0.3$, $\Omega_\Lambda = 0.7$, $w_0 =-1$ and $H_0 = 70\;\rm{km}\;s^{-1}\;\rm{Mpc}^{-1}$.

\section{Methods} \label{sec:methods}
\subsection{Models for Eruptive Mass-Loss}\label{subsec:precursormodel}
As the foundation of each of our theoretical models, we implement the eruptive precursor model described in \cite{2022Matsumoto_PrecursorModel}. 
This model calculates the light curve from hydrogen-rich ejecta with mass $M_{ej}$ promptly launched by an eruption from the surface of a progenitor with radius $R_*$. The ejecta are assumed to expand homologously, and its velocity profile is characterized by density through the self-similar solution of \cite{1960sakurai1960_shockedgas}, $v\propto \rho^{-\mu}$, where $\mu=0.22$ (corresponding to the polytropic index of the envelope of $n=3/2$). The ejecta is divided into multiple velocity shells and its thermal evolution is independently solved. The bolometric luminosity is then calculated as the sum of the diffusion luminosity contributed by each shell. The spectrum is approximated by a black body with a temperature determined by the photosphere (the shell whose optical depth $\tau$ is unity). The model is parameterized by the minimum eruption velocity, $v_{ej}$, the total ejected mass, $M_{ej}$, the progenitor radius, $R_{*}$, mass, $M_{*}$, and polytropic index, $n$. This model does not encode a physical mechanism powering the emission, and is able to qualitatively reproduce the behavior of both short-lived ($<$50d) and more long-lived ($>$100d) precursors \citep{2022Matsumoto_PrecursorModel}. 

%The model is parameterized by the minimum eruption velocity, $v_{\rm{ej}}$, the total mass ejected, $M_{\rm{ej}}$, and the progenitor radius, $R_{*}$, mass, $M_{*}$, and polytropic index, $n$. The eruption ejecta are assumed to follow a shock velocity profile of $v\propto \rho^{-\mu}$, where $\mu=0.22$ \citep{1960sakurai1960_shockedgas}, and the material is assumed to be launched from the surface of the star. The observed luminosity from the eruption is driven by a combination of the energy deposited by the shock passage and the recombination photons from the hydrogen-rich ejecta as it is accelerated to beyond the escape velocity. This model is able to qualitatively reproduce the behavior of both short-lived ($<$50d) and more long-lived ($>$100d) precursors \citep{2022Matsumoto_PrecursorModel}.

In this model, the peak luminosity of the observed emission is strongly correlated with the progenitor radius, as material becomes unbound more easily for progenitors with more tenuous envelopes. We assume a RSG progenitor to simulate the dominant contribution to the detected precursor population relative to more compact progenitors such as Blue Supergiants (BSGs). We fix $n=3/2$ and $M_{*} = 10\;M_{\odot}$, and generate a model grid of 8000 spectral energy distributions (SEDs) spanning [1000, 15000]$\;\mathrm{\AA}$ and [0, 200]$\;$d relative to eruption. For each SED, we sample $R_*$ values in 20 log-uniform bins spanning $[100, 1000]\;R_\odot$, $v_{ej}$ in 20 linear-uniform bins spanning $[0.1, 1]\times\;10^3\;\rm{km}\;s^{-1}$, and $M_{ej}$ in 20 linear-uniform bins spanning $[0.01, 1]\;M_{\odot}$. We refer to this model grid as `MM22/II', where the `II' designates a precursor model for SNe~IIP/IIL (in contrast to precursors for SNe~IIn, which we describe in additional detail below)\footnote{\cite{2022Matsumoto_PrecursorModel} also explore pre-explosion mass-loss due to enhanced winds, but the implementation of this model is beyond the scope of this work.}.

An alternative eruption model developed by \cite{2021Linial_Precursors} considers material shed from a polytrope and accelerated to the progenitor's escape velocity by a Sedov-like blast wave \citep{1946Sedov_ShockWave} launched from the stellar core (e.g., following a change in nuclear core burning). Through a series of hydrodynamical simulations, \cite{2021Linial_Precursors} establishes a physically-motivated range of ejecta masses associated with eruptive precursors to RSGs. This ejected mass range extends lower than the values considered in \cite{2022Matsumoto_PrecursorModel}, and is determined by the location at which the outward-propagating shock wave transitions from a slow to a fast shock. Because the energy deposition occurs deeper into the star than the \cite{2022Matsumoto_PrecursorModel} model (where the material is launched directly at the surface), less mass is accelerated beyond the escape velocity to become unbound.

\begin{table}[h]
\centering
\begin{tabular}{c|c|c|c}
\hline
Model & Parameter & Unit & Range/Value \\ \hline \hline
MM22/II,IIn & $R_{*}$  & $R_{\odot}$ & [100, 1000] \\ 
       & $M_{ej}$  & $M_{\odot}$  & [0.01, 1] \\ 
       & $v_{ej}$ & $10^{3}$~km$\;$s$^{-1}$ & [0.1, 1]  \\ \hline
L21/II & $R_{*}$  & $R_{\odot}$ & [100, 1000] \\ 
       & $M_{ej}$ & $M_{\odot}$  & [$10^{-5}$, 0.1] \\ 
       & $v_{ej}$  &  & $v_{esc}$ \\ 
\hline
\end{tabular}
\caption{The parameters sampled to construct the three theoretical models for eruptive SN precursors. The MM22/II and MM22/IIn models are constructed from the same SED grids, but with different selection cuts and priors on $R_*$ (see text for details).}
\label{tab:models}
\end{table}

To explore a range of possible precursor events, we define a new model grid in which we sample ejecta masses within $M_{ej}\;\epsilon\;[10^{-5}, 10^{-1}]\;M_{\odot}$, the range considered in \cite{2021Linial_Precursors}, and self-consistently update the minimum velocity of each precursor $v_{ej}$ to the escape velocity of the progenitor. We refer to this model grid as `L21/II'. We summarize the values of the sampled parameters in our MM22/II and L21/II model grids in Table~\ref{tab:models}.

%At the lower end, the injected energy must be high enough to launch a shock wave that transitions from slow to fast; if this condition is not met, the shock energy is never deposited in the star and no material is unbound. 

\cite{2021Linial_Precursors} also require that the injected energy is lower than the binding energy of the system, or else no subsequent SN could occur. We apply a similar cut, and eliminate SEDs from our MM22/II grid for which the initial kinetic energy of the precursor exceeds the binding energy of the system (this condition holds for all L21/II SEDs, so none are removed). This leaves \NgridMatsumoto{} SEDs in the MM22/II grid. Imposing this cut forces the assumption for both models that a precursor is caused by energy injected at the center of the star, although because the deposited energy at the stellar surface is always lower than the initial energy at the core due to radiative losses, our upper limit is a conservative one. 
%The M\&M model now reflects the idealized scenario in which the shock wave experiences no radiative losses as it travels to the surface of the star and deposits its energy at the stellar surface. 

The L21/II model assumes the same $v\propto \rho^{-\mu}$ ejecta velocity profile as the MM22/II model. As in \cite{2022Matsumoto_PrecursorModel}, the L21/II SEDs are generated by summing the luminosity contributions of a set of discrete velocity shells in time, solving for the photospheric radius and temperature of the system, and calculating the wavelength-dependent flux of the precursor assuming black-body emission. For the lowest-mass and highest-velocity models in our grid, this results in an early shock-breakout-like peak as the ionized hydrogen from the highest-velocity shell recombines. Because this phenomenon is specific to the assumed mass profile of the ejecta, we manually remove the early peak from each light curve in both the L21/II and MM22/II grids so as not to positively bias the detection statistics. 

Precursor activity has been discovered in only a single normal SN~II, 2020tlf \citep{2022Galan_FinalMoments}. Deep PS1 photometry showed enhanced, persistent emission for 130 days prior to explosion. Although the general timescale of this emission can be reproduced by the MM/II model, the early peak of the model leads to discovery statistics inconsistent with its observed counterpart. To further improve our predictions, we translate the reported bolometric luminosities of the SN~2020tlf precursor in \cite{2022Galan_FinalMoments} into a third SED model, assuming black-body emission. We call this precursor model `2020tlf-like'.

Significantly more precursors have been discovered preceding SNe~IIn \citep{2021Strotjohann_MonthsLong}, in some cases reaching peak luminosities comparable to the lowest-luminosity CCSNe \cite[e.g., the SNe~IIP 2005cs and 2005ay, which exhibited plateaus with $M_V > -16$;][]{2006Tsvetkov_2005cs}. Although the physical mechanism underlying these precursors is unknown, the most luminous SN IIn precursors suggest energies likely to unbind an RSG if deposited at the core. Recently, \cite{2024Tsuna_BinaryOutbursts} proposed a model by which interaction from a compact companion can inject additional energy into the outer envelope of an RSG, allowing precursors to exceed this limit. We further develop a grid of SN~IIn precursor SEDs by considering the full 8000 \cite{2022Matsumoto_PrecursorModel} SEDs (before imposing a precursor energy cut) and imposing a prior on $R_*$ from binary stellar models in MESA (discussed in the following section). This forms our fourth precursor model grid, which we call `MM22/IIn'. 

%* Adding in the CCSN priors
Since the luminosities of a precursor depend in part on the radius of the progenitor launching them, we employ a prior on $R_{*}$ such that our models reflect a physically-motivated underlying stellar population. We discuss these priors in the following section.

%The model grids we have outlined above are valuable for exploring the dependence of the selection function of upcoming surveys on model parameters. They are less useful for exploring the physical properties of the detected precursor population as a whole. We can increase the realism of our intrinsic precursor population by imposing a physically-motivated prior on the progenitor radius $R_{*}$ such that our SEDs are not drawn randomly from each model grid. 
\subsection{Constructing Priors for the Progenitor Radii of Simulated Precursors}\label{subsec:radiuspriors}

We impose empirical priors on the distribution of CCSN progenitor radii with a separate suite of single and binary stellar evolution models (Gilkis \& Laplace et al., in prep.) using version 15140 of the Modules for Experiments in Stellar Astrophysics \cite[MESA;][]{Paxton2011, Paxton2013, Paxton2015, Paxton2018, Paxton2019} code\footnote{\textcolor{black}{All MESA runs can be downloaded via Zenodo:\newline \dataset[10.5281/zenodo.14047988]{\doi{10.5281/zenodo.14047988}}.}
}. We assume a mixing length parameter of $\alpha_MLT = 2$ for all simulations. In superadiabatic regions, we employ the MLT++ method \citep{Paxton2013} \textcolor{black}{at late evolutionary stages (after carbon depletion)} for models with log$_{10}(T_{\rm{eff}}/K)<4.3$, or log$_{10}(T_{\rm{eff}}/K)<5.025$ and log$_{10}(L/L_{\odot})<5.65$
at core carbon depletion. In all other models, the \verb|use_superad_reduction| option is used. 

Simulations were run for three different assumed metallicities: that of the SMC ($Z=0.00224$), the LMC ($Z=0.0056$), and the Milky Way ($Z=0.014$). For the binary simulations, the initial conditions were 27 initial primary masses between 4 and $99\,M_{\odot}$, 3 mass ratios (0.25, 0.55 and 0.85) and 14 initial orbital periods evenly distributed in log-space between 2 and 2223\,d. For single stars, 92 initial masses were used, including those of the primaries in binaries. The explosibility in simulations that reached iron core collapse was determined following \cite{Muller2016}, and in simulations that reached the end of core carbon burning but not iron core collapse an explosion was assumed if the CO-core mass was above the Chandrasekhar mass and below $9\,M_{\odot}$. In total, 1534 simulations were considered to represent SN progenitors: 109 single stars, 1031 primaries in binaries, and 394 merged stars resulting from coalescence during common envelope evolution following unstable mass transfer. Models with hydrogen mass greater than $0.033\,M_{\odot}$ at the end of the simulation were taken to represent a population of SN~II progenitors \citep{Hachinger2012,Gilkis2022}.

We find negligible differences in the radius distributions of the three metallicities, and select the Milky Way model set as our CCSN progenitor radius distribution. We use the \verb|stats| module in \verb|Scipy| to compute a kernel density estimate (KDE) of $R_*$, and sample our KDE with 301 log-uniform bins in the range $\rm{log}(R_*/R_{\odot})\;\epsilon\;[2,4]$. In our subsequent LSST simulations, we re-weight the flat probability density function of our precursor SEDs (for both MM22/II and L21/II) by this distribution. For our MM22/IIn models, we adopt the distribution of progenitor radii of all stars in binary systems as our prior on $R_*$. To prevent adding an extra dimension to our precursor model grids, we do not weight our SEDs by the distribution of progenitor masses from these simulations; our detection statistics are only weakly dependent on $M_*$.

%Avishai: previous version of the above two paragraphs commented out:

%We impose empirical priors on the distribution of core-collapse SN progenitor radius with a suite of single and binary-stellar evolution models from the Modules for Experiments in Stellar Astrophysics \cite[MESA;][]{Paxton2011, Paxton2013, Paxton2015, Paxton2018, Paxton2019} code.

%Each model is defined by the initial mass and metallicity of the primary, and the initial mass ratio and period in the case of a binary system. Some of the models begin as binaries but coalesce in the simulation during common-envelope evolution. The calculated CO-core mass for the models that reach iron core-collapse by the end of the simulation are used to predict the progenitors that will explode as SNe, and the models with hydrogen mass greater than $0.033\;M_{\odot}$ at the end of the simulation are taken to be SN~II progenitors (Laplace et al., in prep.). Simulations are run for three different assumed metallicities: that of the SMC ($Z=0.00224$), the LMC ($Z=0.0056$), and the Milky Way ($Z=0.014$). We find negligible differences in the radius distributions of these three cases, and select the Milky Way set as our CCSN progenitor radius distribution. We use the \verb|stats| module in \verb|Scipy| to compute a kernel density estimate (KDE) of $R_*$, and sample our KDE with 301 log-uniform bins in the range $\rm{log}(R_*/R_{\odot})\epsilon [1,4]$. In our subsequent LSST simulations, we re-weight the flat probability density function of our precursor SEDs (for both MM22/II and L21/II) by this distribution.

\begin{figure}
    \centering
    \includegraphics[width=\linewidth]{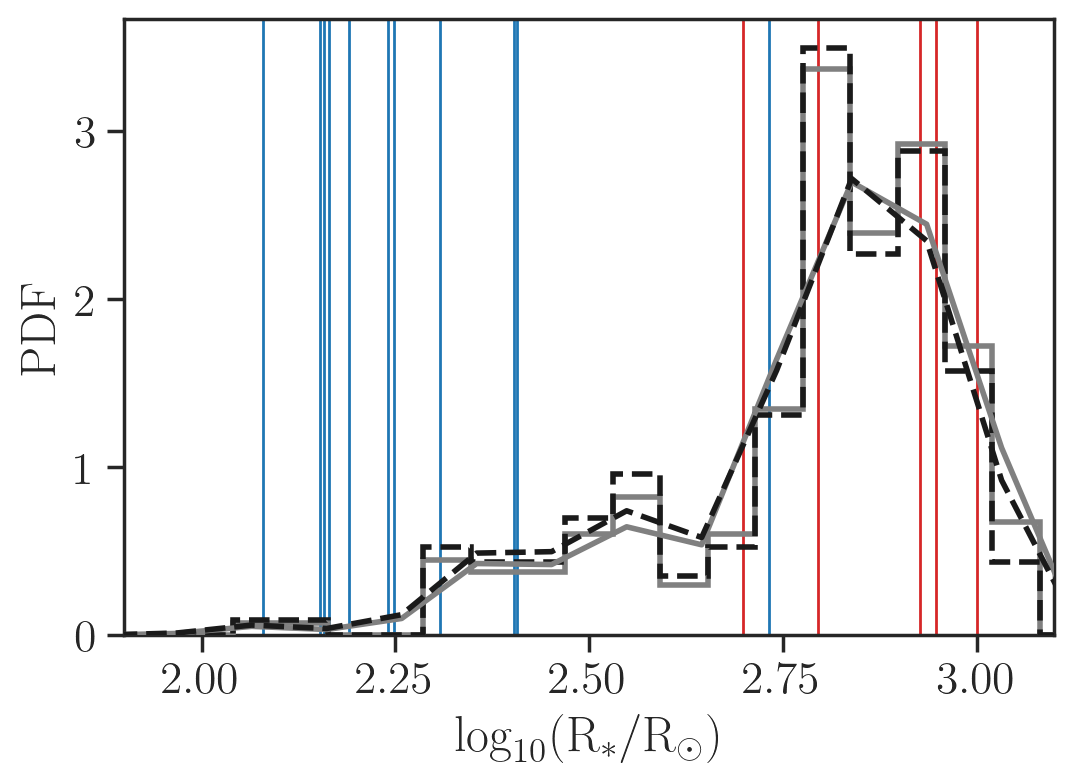}
    \caption{Progenitor radius priors for our IIP/IIL (solid gray line) and IIn (black dashed line) precursor SED grids, derived from MESA simulations of single and binary-star systems evolved to core-collapse. SN~II progenitors are selected by a minimum threshold on hydrogen mass at the end of the simulation, and the radii of candidate SN~II progenitors in binary systems are used to construct the SN~IIn prior. Gaussian kernel density estimates \textcolor{black}{(overplotted smoothed curves)} are made from each sample and used to weight the associated SEDs in the simulation. The inferred progenitor radii of SNe~IIP 2004A, 2009ib, 2017eaw, 2017gmr \citep{2020Goldberg_IIP}, and 2024ggi \citep{2024Xiang_24ggi} are shown as vertical red lines, and those inferred for Galactic Luminous Blue Variables from \cite{2022Mahy_LBVs} using interferometry are shown as vertical blue lines.} \label{fig:radiusPriors}
\end{figure}

We plot the priors on progenitor radii for our IIP/IIL and IIn precursors in Figure~\ref{fig:radiusPriors}. We find a clear overdensity near $R_*\sim10^{3}\;R_{\odot}$, and a slight overdensity at higher radii for binary CCSN progenitors relative to the full population. For comparison, we show the reported progenitor radii for the Type II SN~2004A, SN~2009ib, SN~2017eaw, SN~2017gmr \citep{2020Goldberg_IIP}, and SN~2024ggi \citep{2024Xiang_24ggi}. Despite the small number of events, we find broad agreement with the SN~II progenitor estimates from literature. We caution that our MESA models have assumed hydrostatic equilibrium, and a violation of this assumption in the case of e.g., radial pulsations will lead to variations in the adopted radius distribution.

We also show the radius estimates for a sample of Galactic LBVs from \cite{2022Mahy_LBVs} in Figure~\ref{fig:radiusPriors}. The majority of estimates fall toward the low-density tail of the binary system distribution, inconsistent with the direct detection of LBV progenitors in the case of the SNe~IIn SN~2005gl \citep{2009Nature_2005gl} and SN~2009ip \citep{2010Smith_2009ip,2011Foley_2009ip}. \textcolor{black}{We caution that, at the luminosities of these LBV-like IIn progenitors, our MESA simulations predict direct collapse into a black hole with no associated SN. Late-stage eruptive mass-loss may alter the core structure of these progenitors, altering their explosibility and our subsequent progenitor radius priors \citep[e.g.,][]{2019Woosley_HeStars,2021Laplace}.} Nonetheless, as we discuss in subsequent sections, the resulting model grid is able to reasonably reproduce the properties of observed SN~IIn precursors without excessive fine-tuning of the sampled precursor energies. We note that, for a fixed progenitor mass, a decrease in progenitor radius will lead to a less luminous precursor (Fig.~4 from \citealt{2022Matsumoto_PrecursorModel}), leading to greater inconsistencies between our IIn precursor simulations and those reported in \cite{2021Strotjohann_MonthsLong}. This inconsistency suggests that additional interaction \textcolor{black}{(of a binary progenitor or between multiple shells/clumps of surrounding CSM)} contributes to the observed emission of precursors, as will be discussed later.

% It remains unknown whether binary interaction contributes to the energy budget of IIn precursors;
%and the original MM22/II model cannot be applied to the binary scenario of \cite{2024Tsuna_BinaryOutbursts}

%We construct an SED for this case using the reported bolometric luminosities and color temperatures for the flat (\textbf{omega value here}) and isotropic (\textbf{omega value here} scenarios, giving each equal weight in our simulation. We deem this model \textbf{T24}. 

\begin{figure*}
    \centering
    \includegraphics[width=\linewidth]{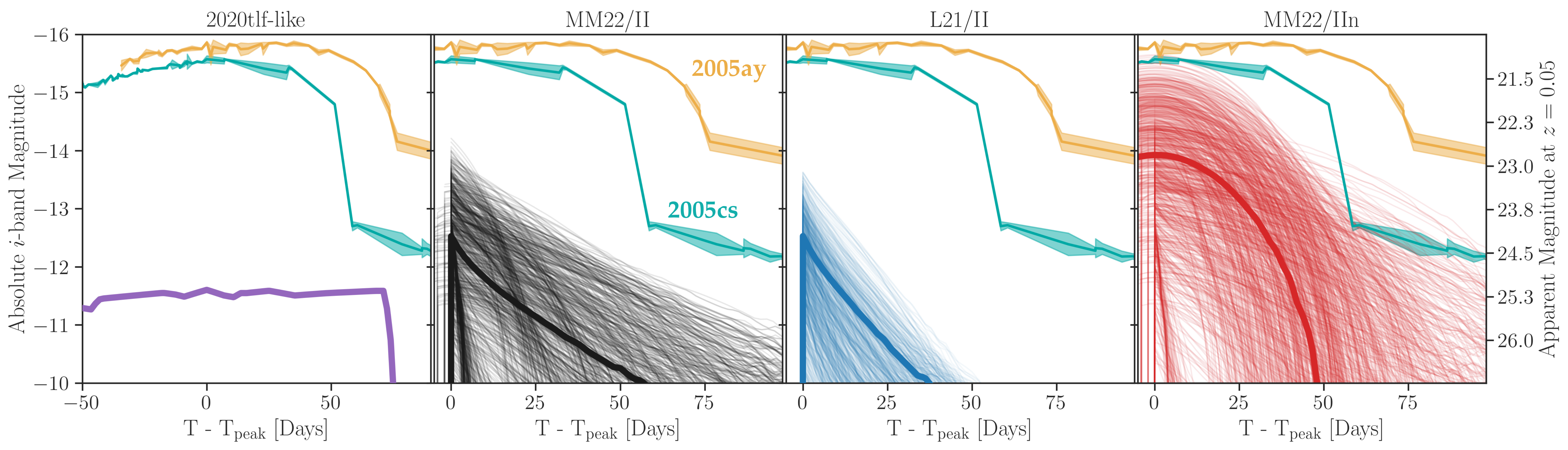}
    \caption{$i$-band light curves for 1000 model SEDs (\textcolor{black}{thin} lines) and a single characteristic event (\textcolor{black}{thick} line) for the four CCSN precursor models considered. \textcolor{black}{Precursors are color-coded by model type: 2020tlf-like (purple), MM22/II (black), L21/II (blue), and MM22/IIn (red).} $I$-band light curves for the extremely underluminous SNe~IIP 2005cs (turquoise) and 2005ay (yellow) are shown for comparison. Phases for all light curves are shown relative to $i$-band peak. Lower typical ejecta masses for the L21/II precursors relative to the MM22/II precursors lead to shorter-lived events. Priors on progenitor radius and additional SEDs above the RSG binding energy lead to higher typical luminosities for the MM22/IIn precursors.}
    \label{fig:ModelLC}
\end{figure*}

We plot high cadence, noise-free light curves in LSST $i$-band for our four precursor scenarios in Figure~\ref{fig:ModelLC}. We show 1000 realizations of each model with low opacity, and a single representative event as a solid line. As expected, the $i$-band brightness of the 2020tlf-like model stays roughly constant for longer than each of the other models. The lower ejecta masses of the L21/II model relative to the MM22/II model manifest in a population of shorter-lived events. The typical brightness of events from the MM22/IIn model grid generally exceeds those of the other three models, and the longest-lived events exceed 75~days. For comparison, we also show $I$-band photometry for the low-luminosity SNe~IIP 2005ay and 2005cs retrieved from the Open Supernova Catalog API\footnote{\url{https://github.com/astrocatalogs/OACAPI}}.

% The highlighted MM22/II, L21/II, and M22/IIn events were constructed from the same model SED with $M_{\rm{ej}}=0.17\;M_{\odot}$ and $R_{*}= 695 R_{\odot}$; the fiducial ejecta velocity of $v_{\rm{ej}}=385\;\rm{km}\;\rm{s}^{-1}$ is shown in MM22/II and MM22/IIn models but was increased to $v_{\rm{ej}}=407\;\rm{km}\;\rm{s}^{-1}$ in the L21/II model to reproduce the scaling relations in \cite{2021Linial_Precursors}, leading to a shorter-lived event. 
 
\subsection{A Forward Model for LSST Observations}\label{subsec:snana}
We use the \texttt{SNANA} simulation code \citep{2009Kessler_SNANA} as a forward model for our synthetic LSST observations. The LSST DESC collaboration\footnote{\url{https://lsstdesc.org/}} has produced cadences associated with existing and upcoming sky surveys that can be used by the simulation \citep[OpSim;][]{2020Biswas_OpSim}. Starting from a grid of rest-frame SEDs for a transient model, the \texttt{SNANA} code draws an SED (its selection can be weighted, as is done here for progenitor radius), places the transient at a random sky position, reddens it according to the Galactic extinction along the line-of-sight, and selects a random redshift based on a chosen volumetric rate (our adopted volumetric rates will be discussed in detail below). Synthetic magnitudes are computed using filter transmission curves and a cadence associated with a proposed survey strategy. An observed flux and its associated uncertainty are determined for each synthetic magnitude using the OpSim-generated zero point, sky noise and PSF. Detections from the DESC Difference Imaging Analysis pipeline\footnote{\url{https://github.com/LSSTDESC/dia_pipe}} are based on a computed efficiency-vs-S/N curve for LSST \cite{2019Kessler_PLAsTiCC,2022Sanchez_DIA}, obtained through the injection of point sources in synthetic LSST images from the DC2 sky simulations \citep{2021LSST_DC2}. To minimize spurious detections from Poisson fluctuations, we impose a trigger requirement of two detections separated by more than 30 minutes for discovery.

% To avoid spurious detections from Poisson fluctuations, we impose a trigger requirement of two detections separated by more than  30 minutes. More details of the simulation are in 1903.11756.%Synthetic observations are taken following the input survey strategy, and additional files characterizing the detection efficiency, sky-noise, nightly zeropoint, and detection trigger logic are used to produce a set of observations with realistic uncertainties. 

%The \texttt{SNANA} simulation code was initially developed to model distance biases in cosmological analyses using SNe~Ia \citep{2009Kessler_SDSSII, 2014Betoule_SNANA, 2018Scolnic_Pantheon, 2019Kessler_DESCosmology,2021Vincenzi_DESSN}). The ability to realistically reproduce SN observations, coupled with the rise in photometric surveys, has led to the expansion of \texttt{SNANA} to simulate SNe for a series of photometric classification challenges, starting with SNPhotCC \citep{2010Kessler_SNphotCC} and expanding substantially to include non-SN transients and variables with the Photometric LSST Astronomical Time-Series Classification Challenge \citep[PLAsTiCC;][]{2018PLAsTiCC_Dataset,2019Kessler_PLAsTiCC,2019Malz_PLAsTiCC,2023Hlozek_PLAsTiCC} and its successor, the Extended Astronomical Time-Series Classification Challenge \citep[ELAsTiCC;][]{2023Lokken_SCOTCH}. This work is the first to incorporate models for SN precursor emission into the code.

We adopt the latest baseline v3.4 LSST survey strategy in our simulations\footnote{\url{https://community.lsst.org/t/release-of-v3-4-simulations/8548}}, which was released in May 2024 and includes the primary Wide-Fast-Deep (WFD) survey, the sub-survey of five deep drilling fields (DDF), a rolling cadence, and a fraction of observational triplets to increase the median cadence of observations in a single photometric filter. Additional details on this strategy can be found in the Rubin Observatory's Survey Cadence Optimizations Committee recommendations document\footnote{\url{https://pstn-055.lsst.io/}}. While the exact survey strategy adopted by the Rubin Observatory is actively evolving, variations at this stage are expected to be minor and are unlikely to substantially alter the results reported here. 

We assume that all IIP/IIL and IIn SNe are preceded by an eruptive precursor, and adopt the same volumetric rates for their precursors. For our IIP/IIL precursor models (2020tlf-like, MM22/II, L21/II), we multiply the volumetric CCSN rate from \cite{2015Strolger_Rates} by 0.87, following the volumetric fraction estimated from the LOSS sample in \cite{2017Shivvers_Rates}. For our MM22/IIn model, we multiply the rate by 0.05 \citep{2017Shivvers_Rates,2023_ZTFIIn}. 

\textcolor{black}{We caution that these rates are likely an overestimate given the results of \cite{2014Ofek_IInPrecursors}, \cite{2021Strotjohann_MonthsLong} and \cite{2024Reguitti_IInPrecursors}, but emphasize that serendipitous precursor discoveries among both SNe~IIn and the SN~II 2020tlf have been driven by targeted searches preceding nearby explosions. Given the limited depth and baseline coverage of current wide-field surveys (ZTF was commissioned in 2018 and is multiple magnitudes shallower than \textcolor{black}{the} Rubin \textcolor{black}{Observatory} in a single pointing), LSST data can be used to probe fainter, earlier precursors than have been found to date\footnote{\textcolor{black}{\cite{2024Reguitti_IInPrecursors} cautions that their discovery of precursor activity in $\sim$30\% of SNe~IIn is an underestimate of the intrinsic rate driven by observational biases.}}. A multi-year archival search among extant wide-field photometric surveys will shed additional light on relative rates, but the tightest constraints will come from \textcolor{black}{the Rubin Observatory} itself.}

For each precursor grid, we run two simulations. In the first, we simulate the number of precursors recovered in single-epoch photometry spanning the first year of LSST operations (MJD 60796 to 61161). We impose an additional selection cut beyond the trigger criterion, and require at least two detections with a signal-to-noise ratio of $\geq5$ in any bands. In the second simulation, we exclude the detection trigger and our selection cut and write out all events occurring within the observing footprint of the simulated survey. We expand our simulation to the first three years of LSST operations in the second simulation (spanning MJD 60796 to 61526) to mitigate variations due to Poisson statistics, which become relevant when examining binned events discovered in DDF fields.

\section{Results} \label{sec:results}
\subsection{Annual Discovery Rates from Single-Visit Observations}\label{subsec:singlevisit}

\begin{figure*}
    \centering
    \includegraphics[width=\linewidth]{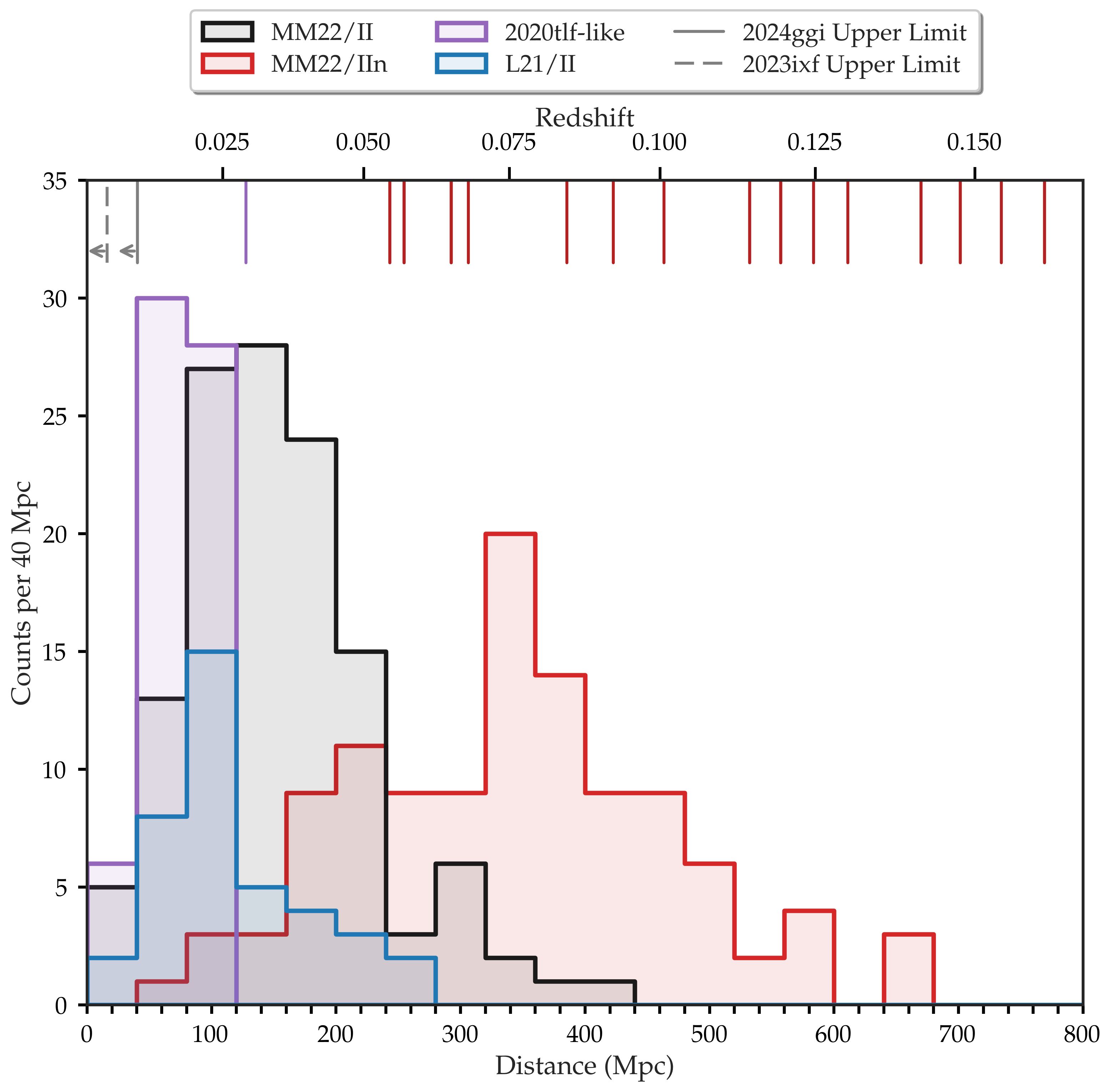}
    \caption{Histogram over distance for the single-visit precursor events passing our two-detection trigger in one year of LSST. The vertical lines at top reflect the distances at which observed precursors to 2020tlf in purple and SNe~IIn in red from \cite{2021Strotjohann_MonthsLong} reach the single-visit limiting magnitude of LSST in $r$. Grey vertical lines indicate the upper limits of precursor emission in 2023ixf \citep{2024Ransome_TwilightYears} and 2024ggi \citep{202Shrestha_2024ggi}.}
    \label{fig:singleVisitKDEs}
\end{figure*}

We first consider the events passing our detection trigger in single-epoch differential photometry (our `single-visit precursors'). We present histograms for the detected event distances of our four models in Figure~\ref{fig:singleVisitKDEs}. We define the median distance $\tilde{d}$ and the 90th percentile of detection distances $d_{90}$ for each model. Due to their low luminosity, 2020tlf-like events are only detected to $\tilde{d}\approx74\;\rm{Mpc}$ and $d_{90}\approx108\;\rm{Mpc}$. Because the distribution for our L21/II models extends toward lower-mass precursors, we observe systematically lower distances for the detected population relative to the \cite{2022Matsumoto_PrecursorModel} grid: we detect L21/II events with $\tilde{d}\approx110\;\rm{Mpc}$ and $d_{90}\approx205\;\rm{Mpc}$, compared to MM22/II events with $\tilde{d}\approx143\;\rm{Mpc}$ and $d_{90}\approx241\;\rm{Mpc}$. The luminous MM22/IIn precursors are detected to significantly greater distances, with over twice the median distance ($\tilde{d}\approx340\;\rm{Mpc}$) and over twice the $d_{90}$ distance of the second-highest model ($d_{90}=497\;\rm{Mpc}$). We summarize these metrics and the total number of detections for all single-visit precursors in Table~\ref{tab:distances}.

\begin{table}[h]
\centering
\hspace*{-15mm}
\begin{tabular}{c|c|c|c}
\hline
Model & $\tilde{d}$ (Mpc) & $d_{90}$ (Mpc) & $N_{\rm{tot}}$ \\ \hline \hline
MM22/II & 143  & 241 & 125 \\ 
L21/II & 110  & 205 & 39 \\ 
2020tlf-like & 74  & 108 & 64 \\ 
MM22/IIn & 341  & 497 & 112 \\ 
\hline
\end{tabular}
\caption{Median distance, 90th-percentile distance, and total number of single-visit precursors detected in one year of LSST.}
\label{tab:distances}
\end{table}

For comparison, we also show the maximal LSST detection distances of the SN~2020tlf precursor from \cite{2022Galan_FinalMoments} and observed SN~IIn precursors from \cite{2021Strotjohann_MonthsLong} (using the anticipated LSST limiting magnitude) as vertical lines in Figure~\ref{fig:singleVisitKDEs}. Overall, we find good agreement between our simulated detection distances and the observed events.%, and observe a sharp drop-off in 2020tlf-like event at the 2020tlf detection limit. 
 The observed SN~IIn precursors extend toward the higher end of our predicted distribution, but this is due to the greater abundance of longer-lived events that can be recovered through binning (as is done in \citealt{2021Strotjohann_MonthsLong}). We also show the reported upper limits for emission preceding the nearby Type II SN~2024ggi and SN~2023ixf from \cite{202Shrestha_2024ggi} and \cite{2024Ransome_TwilightYears}, respectively. Neither event showed optical signatures of precursor activity despite their close proximity; we will return to these events in \S\ref{subsec:dust}.

\begin{figure}
    \centering
    \includegraphics[width=\linewidth]{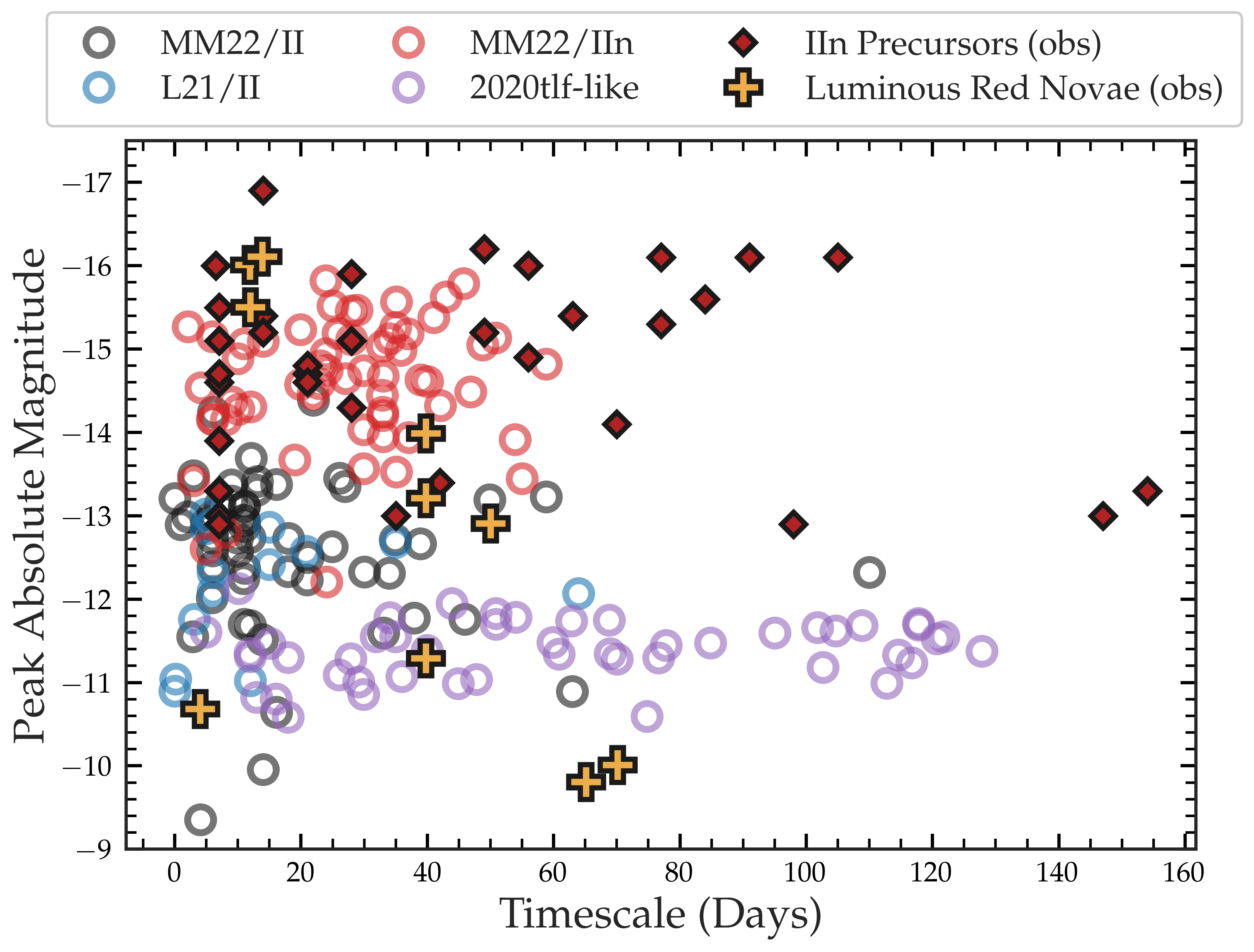}
    \caption{Brightness versus observed timescale for the single-visit precursors passing the detection trigger, where the timescale is defined as the number of days from first to last 5-$\sigma$ detection in any band. Models are listed in the legend. Also shown are the brightness\textcolor{black}{es} and timescales for observed SN~IIn precursors from \cite{2021Strotjohann_MonthsLong}, and those of Luminous Red Novae (LRNe) reported in \cite{2013Kasliwal_BridgingtheGap}.}
    \label{fig:magTimescale}
\end{figure}

Next, we investigate the properties of the discovered events. We plot the peak absolute brightness in any band for the single-visit precursors as a function of the observed timescale, which we have estimated as the number of days from first to last detection in any band. We caution that this observational definition does not directly correspond to the intrinsic timescale of an event, particularly for events poorly characterized at higher redshifts. We present the results in Figure~\ref{fig:magTimescale}, and compare our simulations to the properties of the reported IIn precursors from \cite{2021Strotjohann_MonthsLong}. We also overplot a sample of luminous red novae (LRNe) from \cite{2013Kasliwal_BridgingtheGap}. These transients are believed to be powered by hydrogen recombination following the ejection of a common envelope or merger of a binary stellar system \citep{2019Pastorello_LRNe}, and their similarity to IIn precursors has been discussed extensively in the literature \citep[the \citealt{2022Matsumoto_PrecursorModel} model we employ has also been used to model LRNe;][]{Matsumoto&Metzger2022b}. 

The brightness distribution of synthetic and observed IIn precursors is comparable, but we are unable to reproduce the sub-population of longer-lived ($>$50d) IIn precursors. We note that multiple precursor events spanning $\sim$100 days have been reported for the SNe~IIn in \cite{2021Strotjohann_MonthsLong}. Because these long-lived events will be easier for LSST to detect in both single-visit and binned searches, our subsequent annual rates for SN~IIn precursors can be taken as a conservative estimate. 

We observe significant scatter in the estimated timescale for 2020tlf-like events due to the flat, long-lived emission, which can sit at the LSST detection limit and be marginally detected above the noise in only a few observations. As expected, the observed population of LRNe spans the full parameter space of our precursor models. This makes them a primary contaminant of upcoming precursor searches. %(\citealt{Matsumoto&Metzger2022b} introduced an LRNe model using the same framework as the one for precursors in \citealt{2022Matsumoto_PrecursorModel}). 

\begin{figure}
    \centering
    \includegraphics[width=\linewidth]{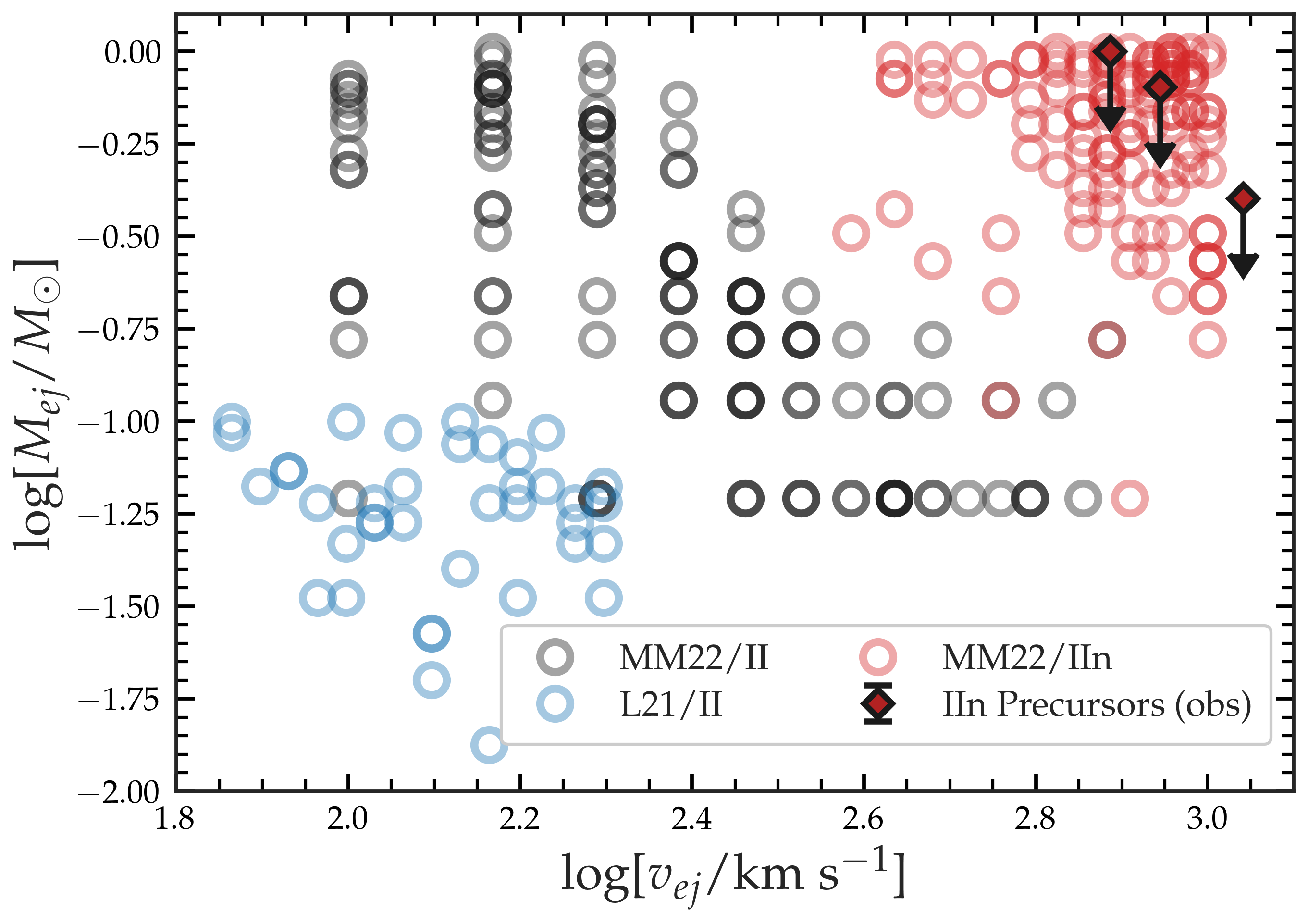}
    \caption{Ejecta masses as a function of the minimum ejecta velocities for the single-visit precursors passing the detection trigger. Precursor models are listed in \textcolor{black}{the} legend. We also show the properties of well-constrained IIn precursors detected in \cite{2021Strotjohann_MonthsLong}: their estimated CSM velocities from narrow line profiles in early SN spectra, and upper limits to the inferred ejecta mass from SN rise times.}
    \label{fig:ParamComparison}
\end{figure}

Next, we consider the distribution of physical parameters for single-visit MM22/II, MM22/IIn, and L21/II precursors. We plot $M_{ej}$ versus $v_{ej}$ for detected events in Fig.~\ref{fig:ParamComparison}, and superimpose the well-constrained velocities and ejecta mass limits reported by \cite{2021Strotjohann_MonthsLong}. The CSM velocities in this work have been estimated from shock-ionization line profiles, and the mass limits are inferred from the maximum allowable diffusion time consistent with the timescale of the subsequent SN rise. Again, we find good agreement between detected and simulated IIn precursors, although we caution that the literature values are order-of-magnitude estimates. We find minimal overlap between parameter properties between models, with L21/II precursors occupying the lowest velocities and ejecta masses. 

%\footnote{the ejecta masses were constrained by comparing the diffusion timescale of the associated ejecta with the rise times of the subsequent SN, and the velocities were inferred from the profiles of narrow lines in the SN spectrum}

\begin{figure}
    \centering
    \includegraphics[width=\linewidth]{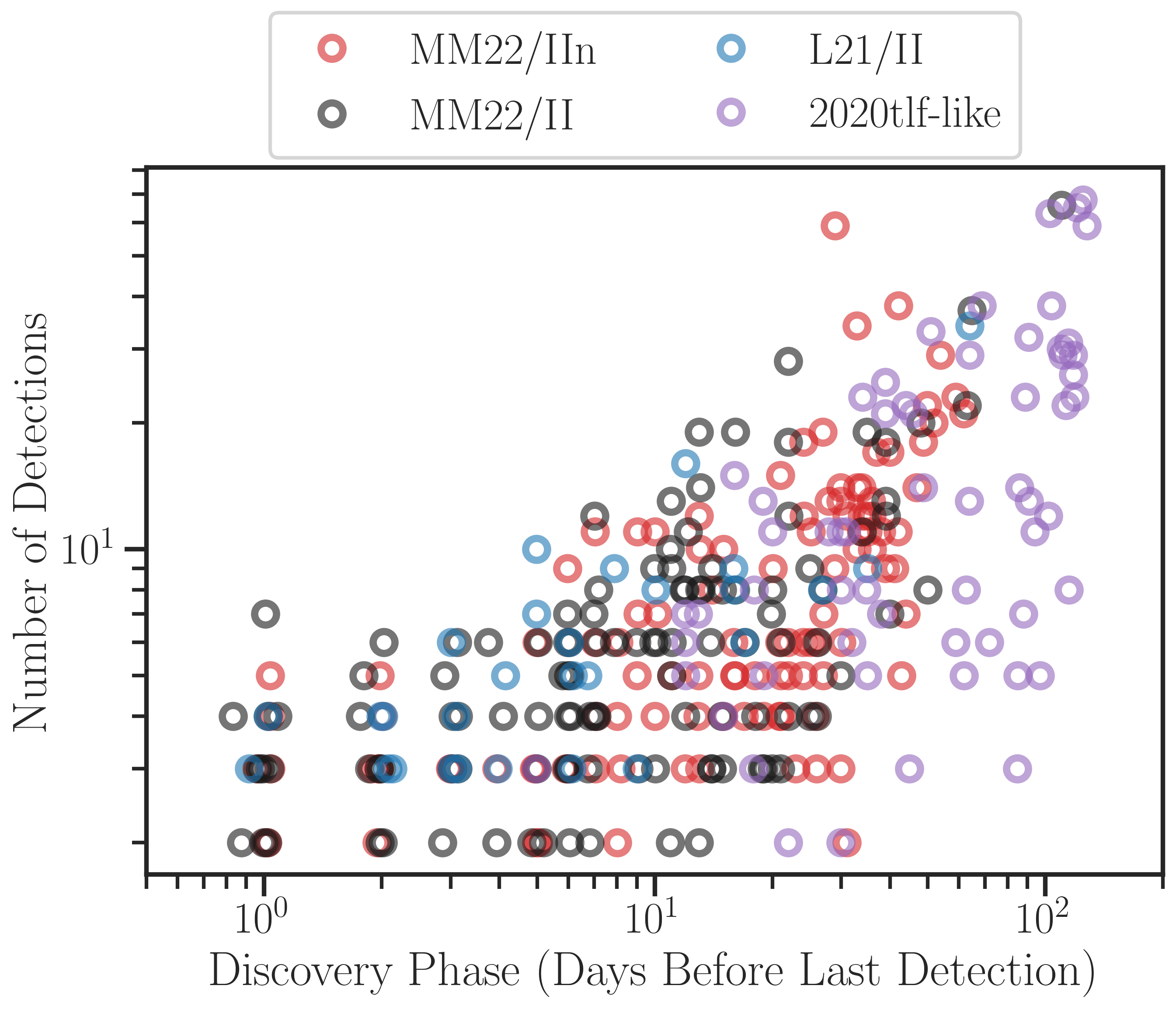}
    \caption{Total number of photometric detections versus the time difference between first and last detection for single-visit precursors. Events top right are ideal for spectroscopic follow-up, while data for events bottom-right are unlikely to be valuable scientifically. Trends in this space are reflective of the the luminosity-timescale relation of each precursor model (Fig.~\ref{fig:magTimescale}).}
    \label{fig:detectionphase}
\end{figure}

In Fig.~\ref{fig:detectionphase}, we compare the total number of detections for each single-visit precursor versus the number of days between first and last detection. Events in the top-right corner of this phase-space are ideal for photometric and spectroscopic follow-up, while follow-up will not be possible for the events in the bottom-left. Again, we find a strong dependency on model type, with 2020tlf-like precursors ideal follow-up targets due to their longer timescales followed closely by MM22/IIn precursors due to their high relative luminosities. 

\subsection{Annual Discovery Rates from Binned Photometry}\label{subsec:binnedvisits}
Discovery from a two-detection trigger may be possible for the brightest tail of the precursor luminosity distribution, but the majority of {\it observed} precursors have been discovered through stacking of archival photometry \textit{a posteriori} once an SN is found. We investigate this possibility by excluding all LSST detection triggers from our observing model and saving the differential photometry for all events of each model simulated within $z<0.3$ and across the first three years of LSST. 

For each event, we bin the differential photometry separately in each LSST filter using a fixed bin size $N$ in days. We adopt a similar detection criterion for our binned photometry as for the unbinned case, and require at least one $S/N\geq\;$5 binned detection. We consider 4 bin sizes: $N=$1, 20, 50, and 100 days. Because the typical single-filter cadence of LSST in Wide-Fast-Deep mode is $\sim$18 days, bin sizes less than $20$ days are unlikely to consistently capture more than a single observation. Further, the vast majority of our precursors do not last longer than $50-75$ days; using a wider bin would average out any potential detections with the background.

\begin{figure*}
    \centering
    \includegraphics[width=\linewidth]{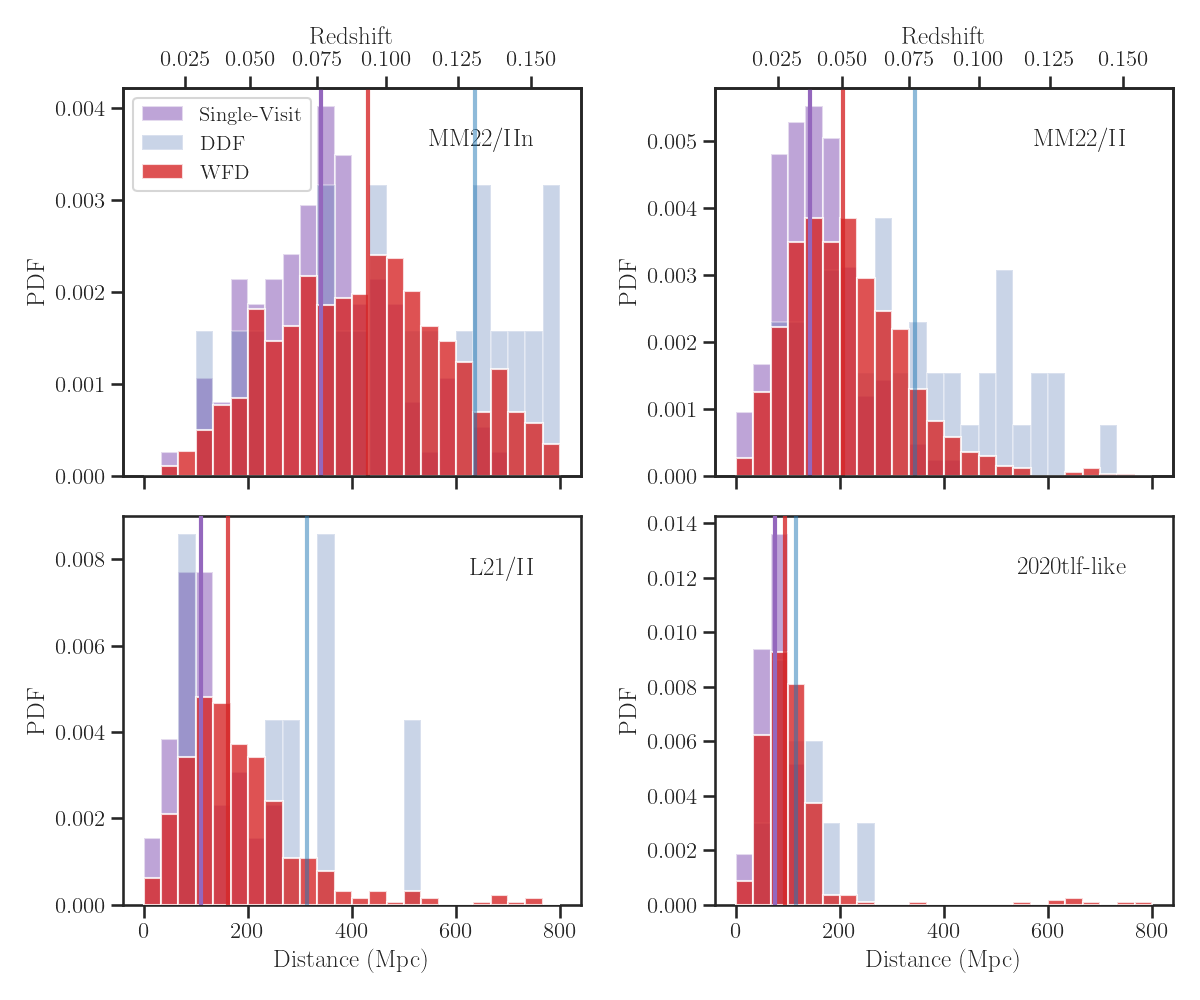}
    \caption{Distances of precursors detected in LSST data in the idealized binning case at any bin width during the first three years of the primary Wide-Fast-Deep survey (WFD, red) and the Deep Drilling Field survey (DDF, blue). For comparison, we also show the distances of precursors from our 1-year single-visit study described in \textsection\ref{subsec:singlevisit} (purple). The median of each distribution is shown as a solid line.}
    \label{fig:WFD_vs_DDF}
\end{figure*}

We present the three-year recovery statistics for our four classes in the WFD and DDF surveys across all bin widths in Fig.~\ref{fig:WFD_vs_DDF}, and overplot the single-visit detections described in the previous section. We find a $\sim$70\% increase in the median detection distance in the three-year WFD photometry relative to single-visit across all models: up to 431~Mpc for the MM22/IIn precursor, 206~Mpc for the MM22/II precursor, 162~Mpc for the L21/II precursor, and 94~Mpc for 2020tlf-like precursors. The WFD distributions all feature a substantial positive skew, where faint precursors are marginally detected above the background. 

As expected, the median detection distance is also higher in the DDF fields than in WFD as a result of the higher single-band cadence. In all SN~II models (MM22/II, L21/II, 2020tlf-like), we nearly double our median recovery distance relative to WFD. These longer-lived, more abundant precursors relative to the MM22/IIn model are significantly aided by the increased DDF cadence. The median distance for recovered MM22/IIn precursors increases by a smaller fraction, from 432 to 637~Mpc. This is a consequence of the relative luminosity of these events and the lower assumed volumetric rate (such that a larger fraction of the intrinsic population is detected with the baseline WFD cadence than the SN~II precursors). %Interestingly, the most distant precursor detected in both DDF and WFD is $\sim$800 for every model class.

\begin{figure*}
    \centering
    \includegraphics[width=1.\linewidth]{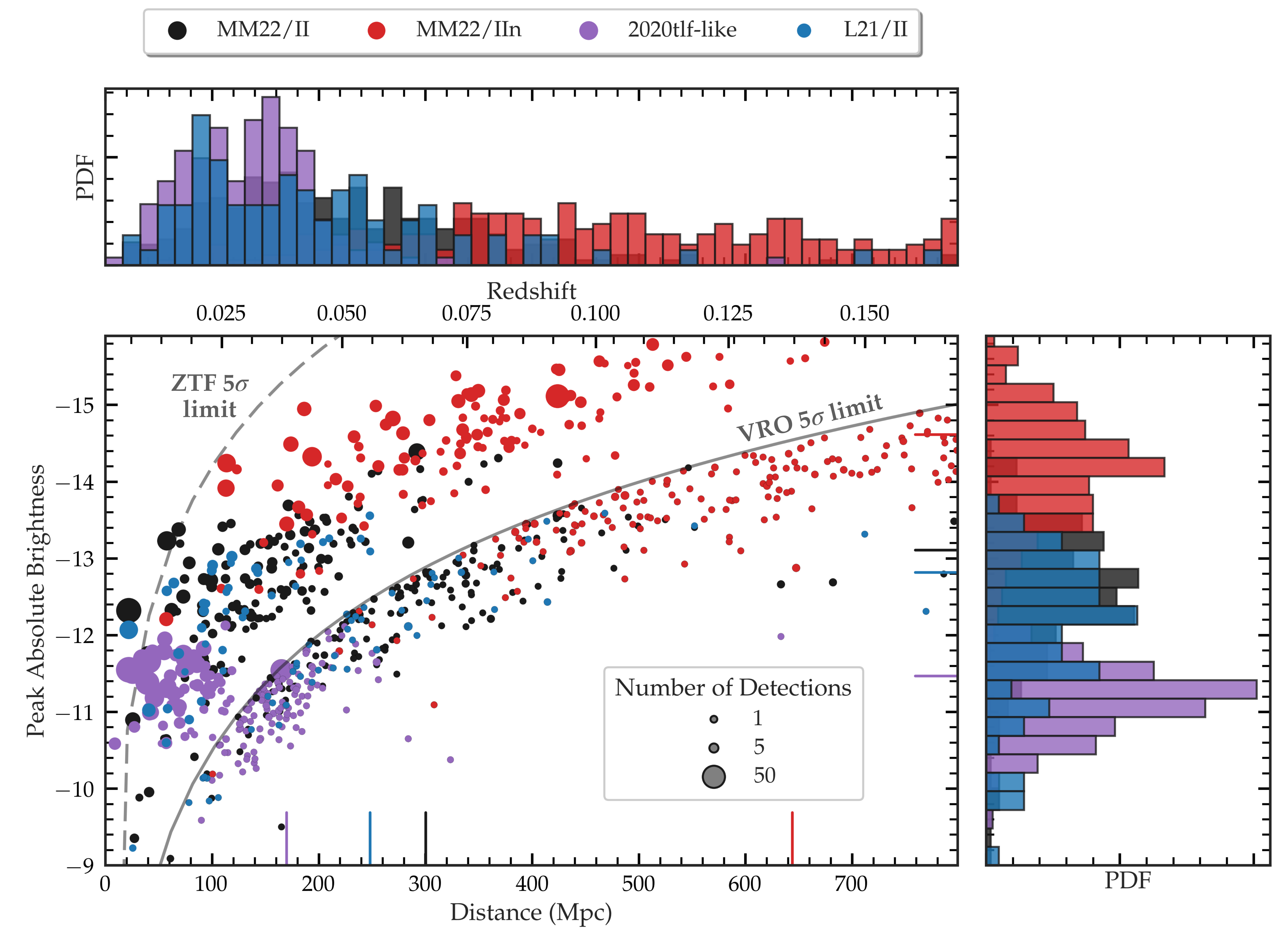}
    \caption{All single-visit and binned CCSN precursors from one year of synthetic LSST observations. Color corresponds to the precursor model assumed (given in legend) and point size corresponds to the total number of 5-$\sigma$ detections for that precursor. Dashed gray line corresponds to the $r$-band single-visit 5-$\sigma$ detection limit of the Zwicky Transient Facility (ZTF) and solid gray line corresponds to the $r$-band single-visit 5-$\sigma$ detection limit of the Rubin Observatory. Binned detections are only shown beyond the single-visit VRO detection limit for clarity. Colored lines at bottom and right give the upper 75th percentile for distance and peak brightness for each model type.}
    \label{fig:detectionVDistance}
\end{figure*}

Finally, we present the peak absolute brightness as a function of distance for all single-visit and binned precursors in Fig.~\ref{fig:detectionVDistance}, and compare these to the 5-$\sigma$ single-visit $r$-band detection limit of LSST and ZTF. We find that the vast majority of detections for all precursor models surpass the ZTF $r$-band single-visit 5-$\sigma$ depth of $\sim$20.8 \citep{2019Masci_ZTF}, reflecting the necessity of the Rubin Observatory to characterize dim transient populations such as the ones considered here. We also find that the events discovered by binning will have very few statistically-significant detections for characterization, and may only be scientifically valuable to constrain demographics (unless timescales are inferred by comparing multiple binning approaches, as we discuss in \textsection\ref{subsub:bin_strategies}).

\textcolor{black}{We can broadly compare our recovery statistics in Fig.~\ref{fig:detectionVDistance} to previously-detected precursors. Given the observational biases associated with targeted binned searches, we limit our comparison to precursors recovered in our single-visit simulation. \cite{2021Strotjohann_MonthsLong} reports (their Figure~2) that precursor emission was securely detected at the 5-$\sigma$ level preceding 14 SN~IIn in single-visit (unbinned) ZTF pointings, versus none in our one-year simulation. In 10 of these SNe, the brightest detected precursor event had a \textit{median} flux of $M<-15$ in either ZTF-$g$ or ZTF-$r$, brighter than the vast majority of MM22/IIn events. Earlier simulations have been similarly unable to reproduce this high-luminosity tail of observed SNe~IIn \citep[e.g.,][]{2014Shiode_WaveDriven}.}

\textcolor{black}{Given the existence of bright, long-duration SN~IIn precursors not captured by our simulation framework, we have compared our annual single-visit detection rates with those from a simpler IIn precursor model consisting of persistent emission for $\sim$100 days and peaking at $M\approx-16$. The relative rarity of these brighter events among the sample analyzed by \cite{2021Strotjohann_MonthsLong} (associated with $\sim$1\% of all SNe~IIn, and $\sim5\%$ of long-lasting precursors $<$90d from explosion; Figures 7 and 8 of \citealt{2021Strotjohann_MonthsLong}) corresponds to a detection rate of $\sim$10 yr~$^{-1}$ with the LSST, sub-dominant to our reported IIn rate. However, we caution that these statistics are highly sensitive to the assumed proportion of this bright sub-population; a relative fraction of 2\% (within 1-$\sigma$ of the fraction reported by \citealt{2021Strotjohann_MonthsLong}) leads instead to a detection rate of $\sim$30 yr~$^{-1}$, with multiple now detected above the ZTF detection limit.}

\textcolor{black}{In 11 of the 14 SNe with single-visit precursor detections, the brightest detected precursor occurred in the final month preceding detonation. CSM velocities far above the values considered here ($v_{ej}>10^{3}\;\rm{km}\;\rm{s}^{-1}$) could increase the luminosity of the observed precursor, but would be inconsistent with values inferred from spectroscopy. Kinetic energy from the interaction between multiple CSM shells ejected prior to the explosion can contribute additional luminosity to an observed precursor, and is suggested by the detection of multiple precursor events preceding several of the SNe~IIn reported in \cite{2021Strotjohann_MonthsLong}. Early eruptions may also alter the density structure of a progenitor envelope, decreasing the photon diffusion time and increasing both the luminosity and duration of subsequent precursor emission \citep{2021Kuriyama_IInMultipleEruptions}. Modeling the long-term response of the progenitor system to these eruptions, and the interaction of CSM from distinct outbursts, may be necessary to reproduce the full diversity of SN~IIn precursors.}

\textcolor{black}{Next, we consider unbinned detections of SN~IIP/IIL precursors. Each of our SN~II models (MM22/II, L21/II, and 2020tlf-like) predict precursor emission detectable in unbinned photometry above the 5-$\sigma$ magnitude limit of ZTF, yet none have been discovered to date. We caution that the timescale and SED of potential SN~IIP/IIL precursors is highly uncertain, informed only by the observational constraints provided by SN~2020tlf (and the single-epoch luminosity and black-body properties of this precursor are not well-constrained). A decrease in the inferred blackbody temperature of 2020tlf-like events of a factor of two pushes the peak of the emission into the infrared and beyond the reach of ZTF (though it may still be detected in LSST-$y$). Furthermore, the persistent emission observed in the 2020tlf precursor may be more consistent with an enhanced wind rather than eruptive mass-loss \citep{2022Matsumoto_PrecursorModel,2022Galan_FinalMoments}. Finally, as we discuss in \textsection\ref{subsec:dust}, extinction from dust surrounding the SN~II progenitor may render these dimmer precursors undetectable at optical wavelengths.}

\textcolor{black}{\subsubsection{Strategies for Binning Pre-Explosion Photometry in the Case of Multiple Precursors}\label{subsub:bin_strategies}}

By default, our simulations have assumed that the templates used to calculate differential photometry contain zero flux. In practice, the selection of a baseline flux level for binning studies is a non-trivial task. Sky noise leads to stochastic variations in flux, host-galaxy light can be substantial for these local events, and a reference flux baseline chosen across some pre-explosion window could be contaminated by marginal emission from the current or a separate undetected precursor event.

To explore strategies to mitigate template contamination by precursor emission, we use the SEDs in Figure~\ref{fig:ModelLC} to calculate the average flux contribution in each LSST passband from each of the precursor models. We consider three strategies for baseline flux estimation:

\textit{I. Baseline Averaging, 50 days:} In the first strategy, we average the flux contribution from the first 50 days of precursor emission, mimicking the pessimistic case in which a precursor is fully captured in the co-added template and the template spans the approximate timescale of precursor emission. We adjust the raw differential fluxes from our three-year simulation according to these new model-specific baselines, and repeat our experiment binning the resulting photometry in 1, 20, 50, and 100-day bins. 

\textit{II. Baseline Averaging, 500 days:} In the second case, we average the flux contribution across 500 days starting at the eruption time of each precursor, in the hopes of averaging out the added flux in the contaminated template. 

\textit{III. Iterative Median:} In the third case, we employ a similar iterative median strategy to the one introduced in \cite{2021Strotjohann_MonthsLong} to find SN~IIn precursors. For each LSST band, we choose a 500-day window starting from eruption time. We calculate the median raw flux across all observations, remove the observation with flux furthest from the baseline (not considering uncertainties), and re-calculate the median flux. We repeat this process until 20\% of the data is left, and select the final median flux as our baseline value in each passband. We then average the contribution across all light curves for each model at a fixed distance. During our binning experiments, we calculate this baseline flux contribution at the distance of our candidate precursor.

\begin{figure*}
    \centering
    \includegraphics[width=\linewidth]{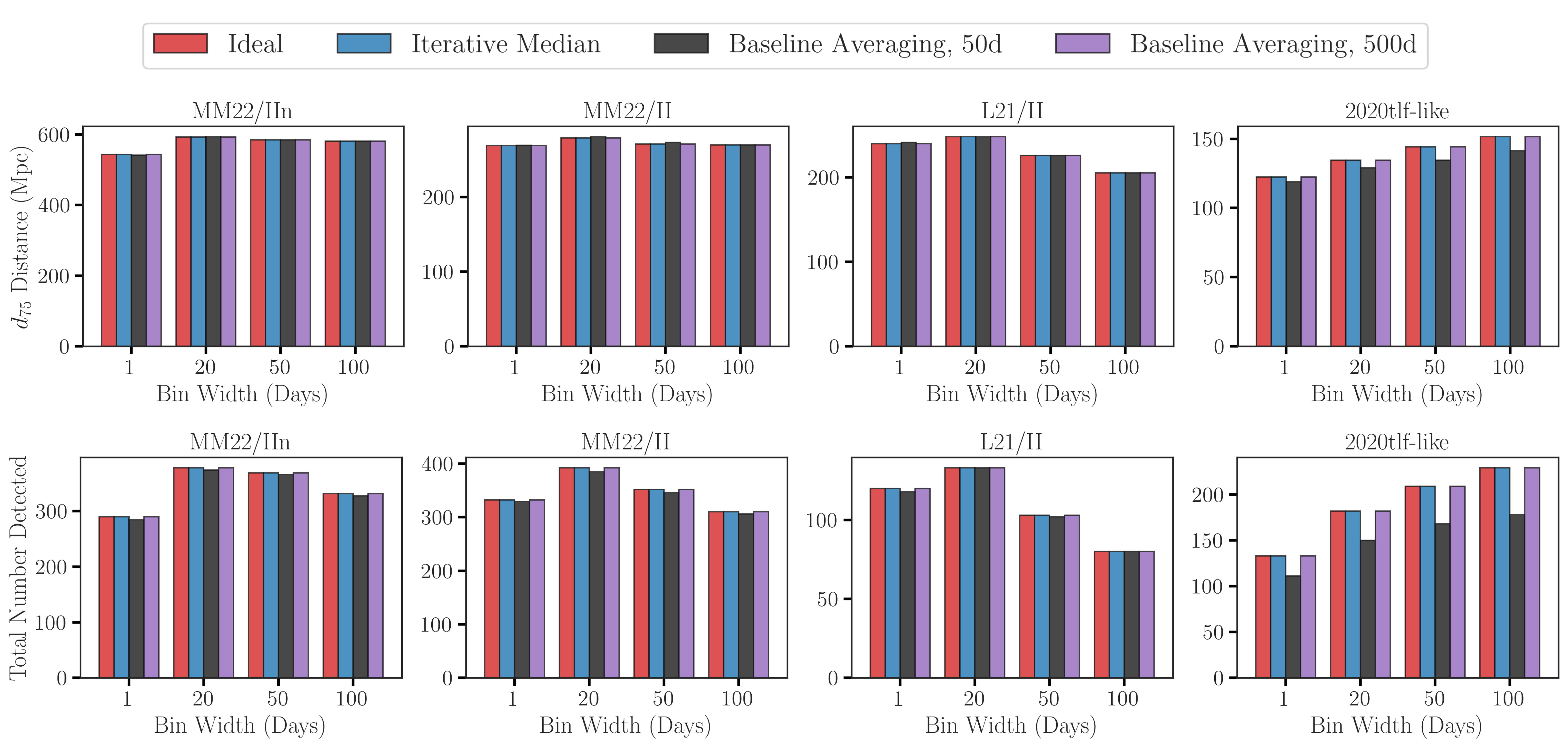}
    \caption{Maximum distance (top) and total number (bottom) of detected precursors in three years of LSST operations as a function of bin width for the four models in this work. Binning with greater than 20 day bins improves the recovery statistics for the long-lived 2020tlf-like precursor due to its long-lived plateau. All others are averaged out in increasing proportion as bin width increases. Events lost from a contaminated template are equally recovered with baseline estimation done via iterative median sampling and by averaging 500d photometry.}
    \label{fig:binnedStatistics}
\end{figure*}

We report the total number of events recovered, and the 70th percentile of their detection distance $d_{75}$, as a function of bin width for each of our three cases (50 and 500 day baseline averaging, and iterative median) compared to our `ideal' uncontaminated baseline, in Fig~\ref{fig:binnedStatistics}. %Because the $d_{90}$ estimate for our three-year study is still dominated by individual observations, we 

The $d_{75}$ distance of recovered precursors is fully determined by the chosen model and bin width, and is not impacted by the baseline strategy adopted. We find the most significant absolute decrease in detection distance across bin widths with the L21/II model, from 240~Mpc with 1-day bins to 200~Mpc with 100-day bins. The MM22/IIn precursors are recovered to comparable distances with every bin width, a result of their high intrinsic luminosities and low volumetric rates relative to the other models.

In the ideal (zero template flux) case, we find an increase in the number of detected events with 20-day binning across all models. With the MM22/II and L21/II models, we observe a decrease in detected events with 50 and 100 day bins, and a similar reduction with 100-day binning for our MM22/IIn model. In contrast, a monotonic increase in the number of detected 2020tlf-like events can be seen increasing the bin width from 1 to 100 days. These differences are reflective of the different phenomenology of each event, and may be useful for broadly characterizing precursor timescales even if individual events are not well-constrained.

When contaminated flux for the 2020tlf-like precursor is averaged across a 50-day window, we lose $\sim$50 precursors at every bin width considered. We recover our full detected population equally by extending our averaging window to 500 days or employing the iterative median strategy. We conclude that our recovery statistics are not significantly impacted by contaminated templates, except in the case of 2020tlf-like precursors. We note that if the flux baseline is systematically increased, a precursor may still be detectable by looking for statistically significant flux \textit{decreases}. These transient events will be flagged in the LSST alert pipeline. 

%We report the total number of precursor events we detect and the distance of the furthest detected precursor for each bin width in Fig.~\ref{fig:binnedStatistics}. We also provide the fractional increase in distance and detected precursors relative to the 1-day bin in the bottom row of Fig.~\ref{fig:binnedStatistics}. Surprisingly, we find that both the highest precursor distance and the number detected \textit{decreases} nearly monotonically for three out of our four models. In contrast, using 100-day bins we find 2020tlf-like precursors to $>$3x the maximum detected distance with 1-day bins. We conclude that the relatively low cadence of LSST in a single filter disfavors a binning approach to increasing our samples, as the precursor models from literature undergo substantial photometric evolution in the gap between observations. With assumptions about the color evolution of the underlying SED, it may be possible to bin across filters, and in this case the 3-4 day cadence may be conducive to discovering far more events; we leave this investigation to future work.

To investigate the impact of Poisson fluctuations on our recovery statistics, we dim the SED of our 2020tlf-like model by 20 magnitudes and run the same binning experiment on all raw flux values at the location of each transient. In our flux-free template, we find no 5$-\sigma$ detections at any bin width, indicating that our precursor estimates are precise. A caveat exists for the 50-day iterative binning technique where the template is contaminated by precursor flux: the technique led to the spurious `identification' of 2-5 events at all bin lengths. We therefore caution that this approach, while able to mitigate the effects of template contamination, may also lead to unreliable statistics for upcoming population studies.

We show sample events discovered by binning in 1-day, 20-day, 50-day and 100-day bins for each of our models in the Appendix. The shaded region indicates the bin in which a detection was made. 

Because these precursors are recombination-driven, the peak of their assumed blackbody emission can be calculated from Wien's law. Assuming the recombination temperature of hydrogen to be $\sim$4000~K, we find a peak of $\sim$7200 \AA, between the effective wavelengths of the LSST-$r$ and LSST-$i$ filters. As expected, the majority of both single-visit and binned precursors at all bin widths are detected in LSST-$r$ and $i$.

\subsection{Maximizing Precursor Identification via a Local Volume Galaxy Survey}\label{subsec:surveyStrategy}

\begin{figure}
    \centering
    \includegraphics[width=\linewidth]{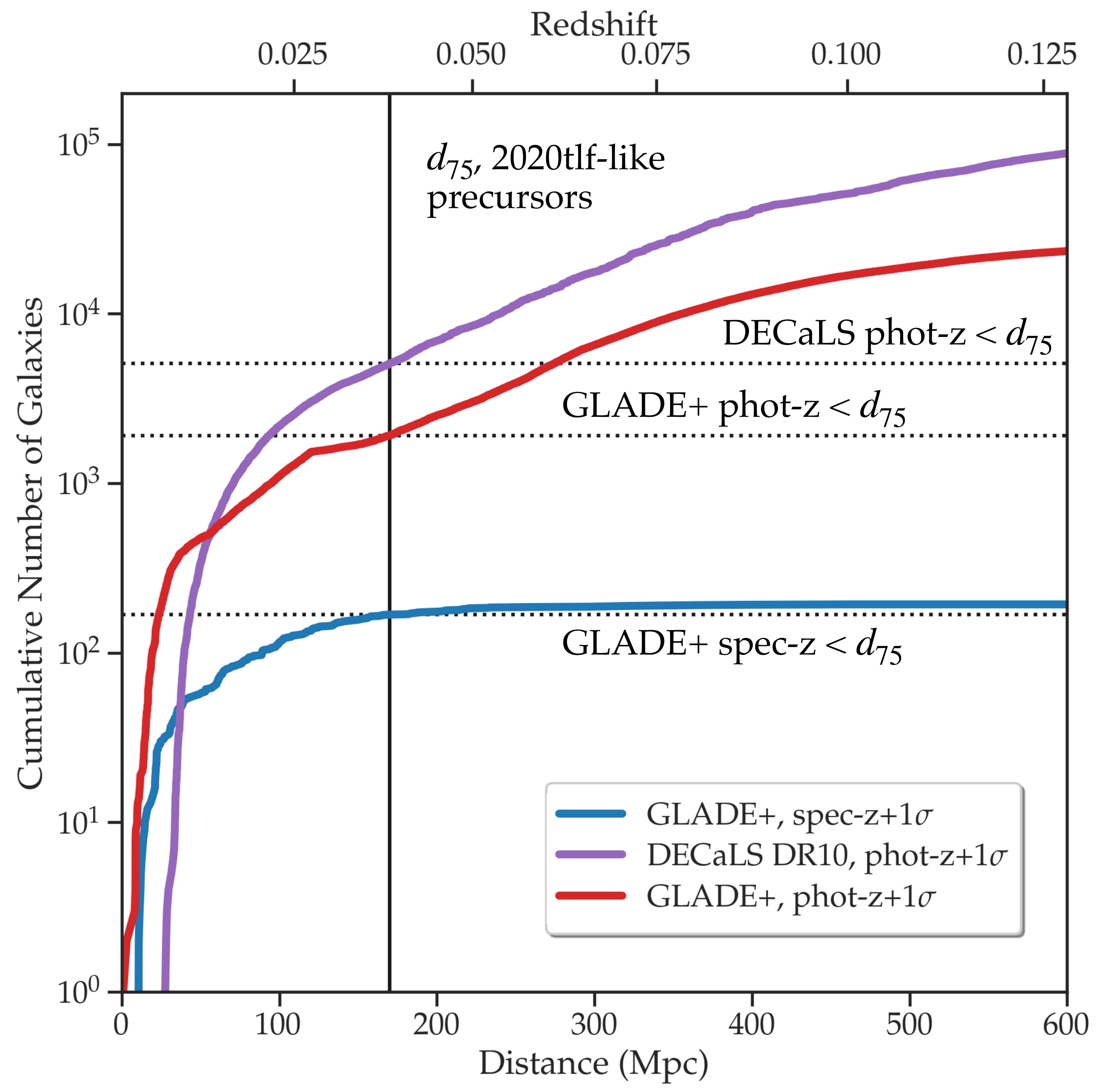}
    \caption{The cumulative number of galaxies in GLADE+ \citep{2022Dalya_GLADE} and DECaLS \citep{2019Dey_DECaLS} overlapping with the LSST survey footprint as a function of distance. We show the spectroscopically-confirmed GLADE+ galaxies in blue, the upper 68th phot-$z$ percentile for the photometric GLADE+ galaxies in red, and the 68th phot-$z$ percentile for the photometric galaxies in DECaLS DR10 in purple. Horizontal lines correspond to the cutoff distance of each catalog within which 75\% of 2020tlf-like single-visit precursors are detected.}
    \label{fig:ngals_surveyvolume}
\end{figure}

SNe detected by LSST will outnumber detected precursors ten thousand to one. Low-latency association of these transients to galaxies at well-constrained distances is essential to clarify their nature as low-luminosity transients. To explore this possibility, we query the GLADE+ \citep{2022Dalya_GLADE} and DECaLS Data Release 10 \citep{2019Dey_DESILegacySurveys} catalogs for all galaxies within 800~Mpc and overlapping with LSST fields. We present the cumulative number in each catalog as a function of distance in Figure~\ref{fig:ngals_surveyvolume}. We overplot the 75th percentile for the distance of detected 2020tlf-like precursors from Figure~\ref{fig:detectionVDistance}. We find 168 spectroscopically-confirmed galaxies within the GLADE+ catalog within this distance, 1912 photometric galaxies within the GLADE+ catalog, and 5090 photometric galaxies within the DECaLS DR10 catalog. Spectroscopically confirming the distances of the photometric galaxies, e.g., through The Dark Energy Spectroscopic Instrument's upcoming MOST Hosts Survey \citep{2024Soumagnac_MOST}, is critical for distinguishing new precursors from the transient zoo and further uncovering the diversity of pre-explosion variability.

The majority of undetected events in our simulation were missed not because of the LSST survey strategy, but because of their intrinsic faintness. A higher-cadence search with smaller-aperture telescopes is unlikely to lead to additional discoveries. As a case study, we take the La Silla Schmidt Southern Survey (LS4) with its forecasted limiting magnitude of $\sim$21\footnote{\url{https://www.snowmass21.org/docs/files/summaries/CF/SNOWMASS21-CF6_CF4_Peter_Nugent-171.pdf}}. A co-added image of 20 background-limited LS4 exposures taken in the span between LSST observations (LS4 will observe 5000 deg$^2$ in alternating nights between $gi$ and $iz$, allowing for 20 $i$-band exposures in 20 days) will reach an apparent magnitude of $\sim22.6$, still magnitudes shallower than the LSST single-visit limiting magnitude of $\sim$24.5. BlackGEM, a wide-field Southern-Hemisphere optical imager operating concurrently to LSST and with an anticipated limiting magnitude of $\sim$23 and a planned Local Transient Survey scanning local overdensities with 6 exposures per night, has greater potential for detecting additional precursors\footnote{\url{https://astro.ru.nl/blackgem/?page_id=302}}. Nonetheless, photometry from smaller instruments will still be useful for \textit{characterizing} events discovered in sparse LSST data. Our single-visit simulations recovered between 12-40 total precursors per year brighter than the co-added LS4 limit. Further, these surveys are useful for characterizing the photometric evolution of the most nearby CCSNe that would saturate the LSST CCDs (a transient peaking brighter than -18th magnitude within $\sim$60~Mpc would saturate the CCD within the anticipated 15s exposure time\footnote{\url{https://www.lsst.org/sites/default/files/docs/sciencebook/SB_3.pdf}}), and constraining signatures of SN-CSM interaction. We therefore recommend a local galaxies sub-survey to complement LSST observations and further enable this science case. 

\subsection{Impact of Dust on Precursor Recovery Rates}\label{subsec:dust}
\begin{figure}
    \centering
    \includegraphics[width=\linewidth]{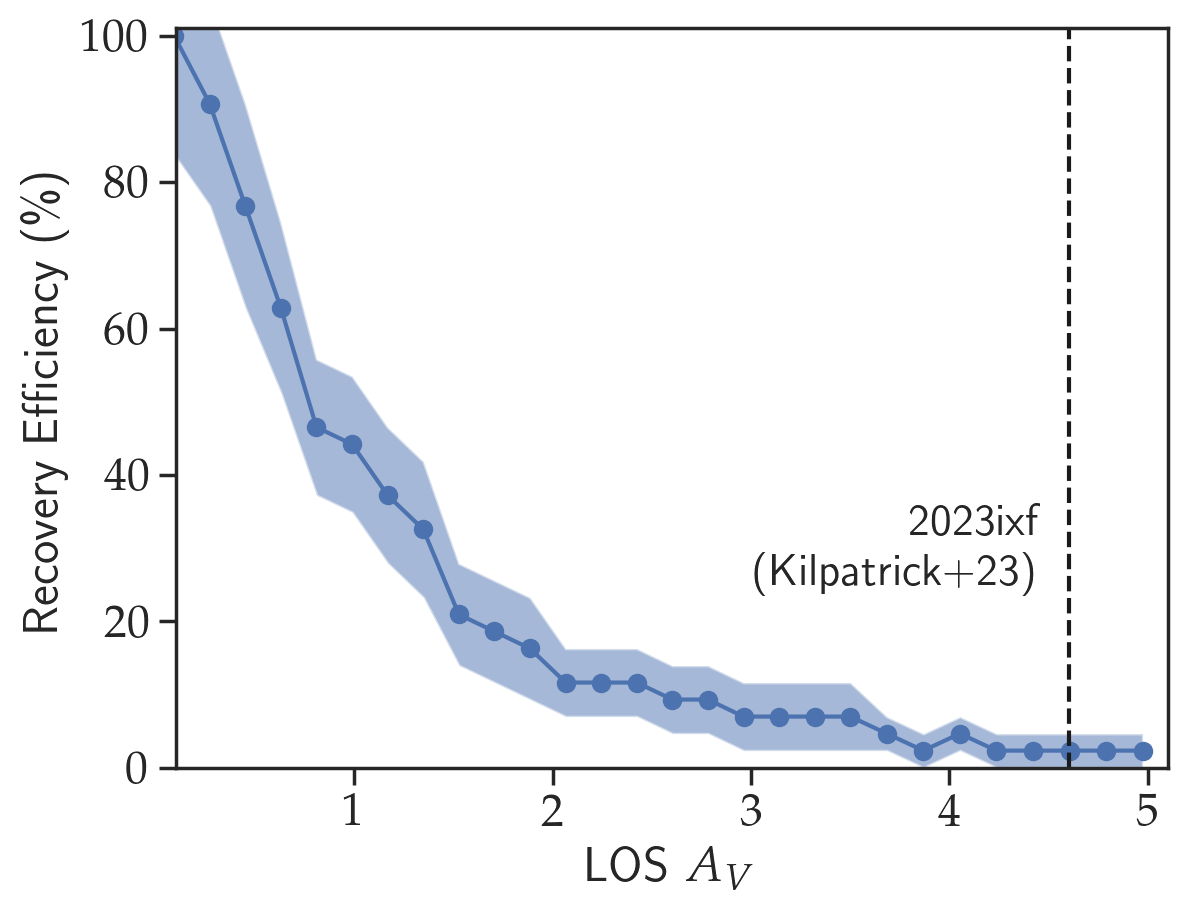}
    \caption{The recovery efficiency of 2020tlf-like precursors as a function of visual-band line-of-sight extinction $A_V$ for a fixed volumetric rate and relative to $A_V=0$. The shaded region gives 1-$\sigma$ uncertainties for each bin, and the black dashed line indicates the reported $A_V$ value for the progenitor of SN~2023ixf from \cite{2023Kilpatrick_2023ixf}.}
    \label{fig:dustRecovery}
\end{figure}

The lack of an optical detection of enhanced pre-explosion emission in the case of SN~2023ixf \citep{2024Ransome_TwilightYears} to an absolute magnitude limit of $\sim -7$, despite years-long infrared variability observed with \textit{Spitzer}, suggests that dust can significantly limit the precursor yield of LSST. Detection limits of $-9.5$ mag were similarly obtained from pre-explosion observations of the SN~2024ggi explosion site, despite flash-ionization lines in early SN spectra suggesting the presence of local CSM \citep{2024JacobsonGalan_2024ggi}. SED modeling of the SN~2023ixf precursor was consistent with a line-of-sight extinction of $A_V\gtrsim 4.6$ mag, where the extinction is dominated by the progenitor environment. \textcolor{black}{While additional limits can be derived from supernova observables, precursor emission can be obscured by substantially more dust destroyed in the subsequent explosion.} To estimate the impact that intervening dust has on precursor recovery rates, we run an additional set of 1-year LSST simulations with the 2020tlf-like precursor model. We assume a fixed volumetric rate and increase the line-of-sight extinction parameter $A_V$ in 27 linearly-spaced bins from 0.1 to 5 mag. We report the recovery fraction in each bin in Fig.~\ref{fig:dustRecovery}. 

The fraction of recovered events drops to 20\%$\pm$7\% at $A_V\approx 1.5$ and 2\%$\pm$2\% at $A_V\approx4.6$ \citep[the reported extinction value for the progenitor of 2023ixf][]{2023Kilpatrick_2023ixf}, dramatically limiting the demographic studies possible with LSST. At intermediate extinction values, a highly reddened precursor may allow some constraints to be placed on dust mass, but large uncertainties on composition and grain size distribution may limit the constraining power of optical photometry alone. The forthcoming \textit{Nancy G. Roman Space Telescope}, which will image 2,000 square degrees of the sky in four NIR filters to a depth of $J \approx 26.7$ AB during its High Latitude Wide Area Survey, will be more naturally suited to detect precursor emission from dust-enshrouded progenitors and reconstruct a warm dust SED. %A joint LSST-\textit{Roman} survey has been proposed \citep{2024Bianco_RomanLSST}, and would permit joint temporal and spectral characterization.

%\textbf{Note about \textit{JWST}?} 

\section{Conclusions} \label{sec:conclusions}

In this work, we have explored the anticipated discovery rates of eruptive precursors to CCSNe with the upcoming Rubin Observatory LSST. Our key findings are summarized below.
\begin{enumerate}
    \item Our models suggest that $40-130$ eruptive precursors to SNe~IIP/IIL, and $\sim 110$ eruptive precursors to SNe~IIn, will be detectable in single-visit LSST photometry annually. This recombination-powered emission will primarily occur in the $r,i$ bands. These rates are strongly dependent on our assumed model: longer-lived \textcolor{black}{and substantially brighter} SN~IIn precursors have been reported in the literature \citep{2021Strotjohann_MonthsLong} than can be explained by the IIn model outlined here, which will further increase detection rates.
    \item Assuming that the template for difference imaging contains zero flux, we anticipate $\sim$400 SN~IIn precursors by binning single-passband LSST photometry in 20-day bins and $\sim$250 2020tlf-like precursors binning in 100 day bins in the first 3 years of LSST. If IIP/IIL precursors are brighter and shorter-lived than the archetypal 2020tlf precursor, as is modeled with the MM22/II eruptions, more may be detectable in the same period.
    \item When the data are binned, the DDF survey is expected to roughly double the median depth at which eruptive precursors to SNe~IIP/IIL are detected relative to the primary WFD survey.
  %  \item When the templates used to select a baseline flux are contaminated by additional precursor emission, the detection of long-lived, 2020tlf-like precursors is most strongly impacted. Increasing the duration of the baseline for estimating the flux, or employing an iterative baseline estimation scheme, can equally mitigate the losses observed. The iterative median technique may introduce additional uncertainties in the events reported. 
    \item Line-of-sight extinction from a dusty CSM can significantly reduce the fraction of recovered precursors with optical photometry, with $\sim$0 events recovered at the reported extinction of the 2023ixf progenitor.
    \item Despite the low number of single-visit events detected relative to other models, 2020tlf-like precursors will be prime targets for spectroscopic follow-up due to their long duration and high number of LSST detections.
    \item Spectroscopic confirmation of precursor host galaxies will be critical to distinguish these rare transients from more distant events, although they will still be easily confused with luminous red novae. Complementary photometric surveys will allow for greater characterization of local CCSNe, offering the possibility of linking detected precursors to the post-explosion behavior of the SN. %Deeper surveys are expected to discover additional precursors. 
\end{enumerate}

%\textit{Facilities:} ZTF \citep{ZTF_image}, ADS, TNS, NED \citep{https://doi.org/10.26132/ned1}, ATel, ANTARES, ALeRCE

\textcolor{black}{Our similar LSST discovery rates between SN~IIP/IIL and SN~IIn precursors at first seem discrepant with observed precursor detections, which are overwhelmingly associated with SNe~IIn. This result is a reflection of the significantly higher volumetric rate of SNe~IIP/IIL and their dimmer modeled precursors, such that a greater fraction of events may be detectable only with \textcolor{black}{the} Rubin \textcolor{black}{Observatory}. Dedicated searches for this hidden population will significantly improve predicted event rates and characteristics.} 

A full treatment of eruptive precursors to CCSNe would model both phenomena simultaneously and self-consistently. Though this approach has been undertaken for individual events \citep[e.g., 2021qqp;][]{2024Hiramatsu_21qqp}, we have avoided imposing assumptions about the phases at which these eruptions occur in this work. Future simulations could include an associated SN using the CSM masses derived from precursor events, and investigate correlations in SN properties. Another extension to this work would be to model the dust produced during precursor episodes and link the line-of-sight extinction to the ejecta mass, although as has been mentioned in \textsection\ref{subsec:dust} this requires imposing additional model assumptions. 

We have considered single-visit and binned precursors separately, but LSST detections of a precursor in the alert stream will prompt binned searches to recover additional observations. This will be possible only for a subset of high-cadence WFD observations for precursors that evolve according to our theoretical models, or for the bulk of observations for longer-lived 2020tlf-like emission. The discovery of an associated SN will aid in selecting specific pre-explosion phases for these more targeted precursor searches.

The number of annual CCSNe within 800 Mpc is low enough that binned searches for pre-explosion emission at every site should be attempted; nonetheless, this represents a major computational undertaking, and support from International Data Access Centers (IDACs) associated with the Rubin Observatory would greatly facilitate this search\footnote{\url{https://www.lsst.org/scientists/in-kind-program/computing-resources}}. Where comprehensive searches are not possible, triaged searches preceding SNe with clear signatures of CSM interaction could be prioritized, although this will severely limit the inferences that could be made about the population at large. 

The number of proposed models for precursor emission continues to grow. Recently, \cite{2024Tsuna_MergerPrecursor} proposed a mechanism by which super-Eddington accretion from a stripped low-mass He primary onto a compact companion drives outflows that power a long-duration transient. The bolometric luminosity of this precursor (with variations from the geometry of the outflows) is constant for years prior to the destruction of the system (via either core-collapse or merger), and brightens to $\sim10^{41}\;\rm{erg}\;\rm{s}^{-1}$ in its final months. This model has been invoked to explain the long-lived precursor emission in the SN~Ibn 20203fyq. While we only consider eruptive mass-loss from hydrogen-rich progenitors in this work, this model will be the focus of a follow-up study. In another recent model, weak shock waves from an RSG interacting with an extended chromosphere are proposed to explain mass-loss rates derived for SN~IIP/IIL progenitors \citep[which are argued to probe winds driven by radiation pressure at the dust formation radius;][]{2024Fuller_BoilOff}. A dearth of optical precursors detected in LSST data would lend support for this theory.

Although we have explored binning strategies when there are multiple eruptions preceding a single SN, we have not explicitly modeled the SED that arises from interacting shells of ejected CSM. A multi-eruption model of this kind was explored in \cite{2023Tsuna_MultiEruption}, and a single-eruption model was also introduced using radiation hydrodynamics. These RSG precursors last hundreds of days, significantly longer than any of the models considered in this work, and peak in the infrared at approximately the luminosity of the 2020tlf-like precursor. Extrapolating the results from our 2020tlf-like results, we expect more optimistic recovery statisics for these events when binning $>$100~day photometry in LSST-$Y$ and higher median distances from events in the Deep Drilling Fields. A systematic comparison between the estimates reported here and the precursors observed in the early years of LSST will shed additional light on the physical mechanisms driving terminal mass-loss in SN~IIn and SN~IIP/IIL progenitors. 

\textit{Software: Astropy \citep{astropy:2013, astropy:2018, astropy:2022}, Matplotlib \citep{Hunter:2007_Matplotlib},  MESA \citep{Paxton2011, Paxton2013, Paxton2015, Paxton2018, Paxton2019}, Numpy \citep{harris2020array}, Pandas \citep{the_pandas_development_team_2024_10957263}, Seaborn \citep{Waskom2021_Seaborn}, Scipy \citep{2020SciPy-NMeth}, SNANA \citep{2009Kessler_SNANA}} 

%%%%%%%%%%%%%%%%%%%%%%%%%%%%%%%%
%%%%%%%%%%%%%%%%%%%%%%%%%%%%%%%%
\section{Acknowledgments} 
\label{sec:acknowledgments}
We thank Itai Linial, Conor Ransome, and Daichi Tsuna for helpful conversations that improved this manuscript. \textcolor{black}{We further acknowledge the anonymous referee for their constructive review.} This work is supported by the National Science Foundation under Cooperative Agreement PHY-2019786 (The NSF AI Institute for Artificial Intelligence and Fundamental Interactions, http://iaifi.org/). 

The Legacy Surveys consist of three individual and complementary projects: the Dark Energy Camera Legacy Survey (DECaLS; Proposal ID \#2014B-0404; PIs: David Schlegel and Arjun Dey), the Beijing-Arizona Sky Survey (BASS; NOAO Prop. ID \#2015A-0801; PIs: Zhou Xu and Xiaohui Fan), and the Mayall z-band Legacy Survey (MzLS; Prop. ID \#2016A-0453; PI: Arjun Dey). DECaLS, BASS and MzLS together include data obtained, respectively, at the Blanco telescope, Cerro Tololo Inter-American Observatory, NSF's NOIRLab; the Bok telescope, Steward Observatory, University of Arizona; and the Mayall telescope, Kitt Peak National Observatory, NOIRLab. Pipeline processing and analyses of the data were supported by NOIRLab and the Lawrence Berkeley National Laboratory (LBNL). The Legacy Surveys project is honored to be permitted to conduct astronomical research on Iolkam Du'ag (Kitt Peak), a mountain with particular significance to the Tohono O'odham Nation.

NOIRLab is operated by the Association of Universities for Research in Astronomy (AURA) under a cooperative agreement with the National Science Foundation. LBNL is managed by the Regents of the University of California under contract to the U.S. Department of Energy.

This project used data obtained with the Dark Energy Camera (DECam), which was constructed by the Dark Energy Survey (DES) collaboration. Funding for the DES Projects has been provided by the U.S. Department of Energy, the U.S. National Science Foundation, the Ministry of Science and Education of Spain, the Science and Technology Facilities Council of the United Kingdom, the Higher Education Funding Council for England, the National Center for Supercomputing Applications at the University of Illinois at Urbana-Champaign, the Kavli Institute of Cosmological Physics at the University of Chicago, Center for Cosmology and Astro-Particle Physics at the Ohio State University, the Mitchell Institute for Fundamental Physics and Astronomy at Texas A\&M University, Financiadora de Estudos e Projetos, Fundacao Carlos Chagas Filho de Amparo, Financiadora de Estudos e Projetos, Fundacao Carlos Chagas Filho de Amparo a Pesquisa do Estado do Rio de Janeiro, Conselho Nacional de Desenvolvimento Cientifico e Tecnologico and the Ministerio da Ciencia, Tecnologia e Inovacao, the Deutsche Forschungsgemeinschaft and the Collaborating Institutions in the Dark Energy Survey. The Collaborating Institutions are Argonne National Laboratory, the University of California at Santa Cruz, the University of Cambridge, Centro de Investigaciones Energeticas, Medioambientales y Tecnologicas-Madrid, the University of Chicago, University College London, the DES-Brazil Consortium, the University of Edinburgh, the Eidgenossische Technische Hochschule (ETH) Zurich, Fermi National Accelerator Laboratory, the University of Illinois at Urbana-Champaign, the Institut de Ciencies de l'Espai (IEEC/CSIC), the Institut de Fisica d'Altes Energies, Lawrence Berkeley National Laboratory, the Ludwig Maximilians Universitat Munchen and the associated Excellence Cluster Universe, the University of Michigan, NSF's NOIRLab, the University of Nottingham, the Ohio State University, the University of Pennsylvania, the University of Portsmouth, SLAC National Accelerator Laboratory, Stanford University, the University of Sussex, and Texas A\&M University.

BASS is a key project of the Telescope Access Program (TAP), which has been funded by the National Astronomical Observatories of China, the Chinese Academy of Sciences (the Strategic Priority Research Program ``The Emergence of Cosmological Structures" Grant \# XDB09000000), and the Special Fund for Astronomy from the Ministry of Finance. The BASS is also supported by the External Cooperation Program of Chinese Academy of Sciences (Grant \#114A11KYSB20160057), and Chinese National Natural Science Foundation (Grant \#12120101003, \#11433005).

The Legacy Survey team makes use of data products from the Near-Earth Object Wide-field Infrared Survey Explorer (NEOWISE), which is a project of the Jet Propulsion Laboratory/California Institute of Technology. NEOWISE is funded by the National Aeronautics and Space Administration.

The Photometric Redshifts for the Legacy Surveys (PRLS) catalog used in this paper was produced thanks to funding from the U.S. Department of Energy Office of Science, Office of High Energy Physics via grant DE-SC0007914.

The Legacy Surveys imaging of the DESI footprint is supported by the Director, Office of Science, Office of High Energy Physics of the U.S. Department of Energy under Contract No. DE-AC02-05CH1123, by the National Energy Research Scientific Computing Center, a DOE Office of Science User Facility under the same contract; and by the U.S. National Science Foundation, Division of Astronomical Sciences under Contract No. AST-0950945 to NOAO.

% Added by EL
This work has received funding from the European Research Council (ERC) under the European Union’s Horizon 2020 research and innovation program (Grant agreement No. 945806) and is supported by the Deutsche Forschungsgemeinschaft (DFG, German Research Foundation) under Germany’s Excellence Strategy EXC 2181/1-390900948 (the Heidelberg STRUCTURES Excellence Cluster).

\bibliography{references}{}
\bibliographystyle{aasjournal}

\appendix

%\onecolumn 
\begin{figure*}[!htbp]
\centering
\includegraphics[width=\linewidth]{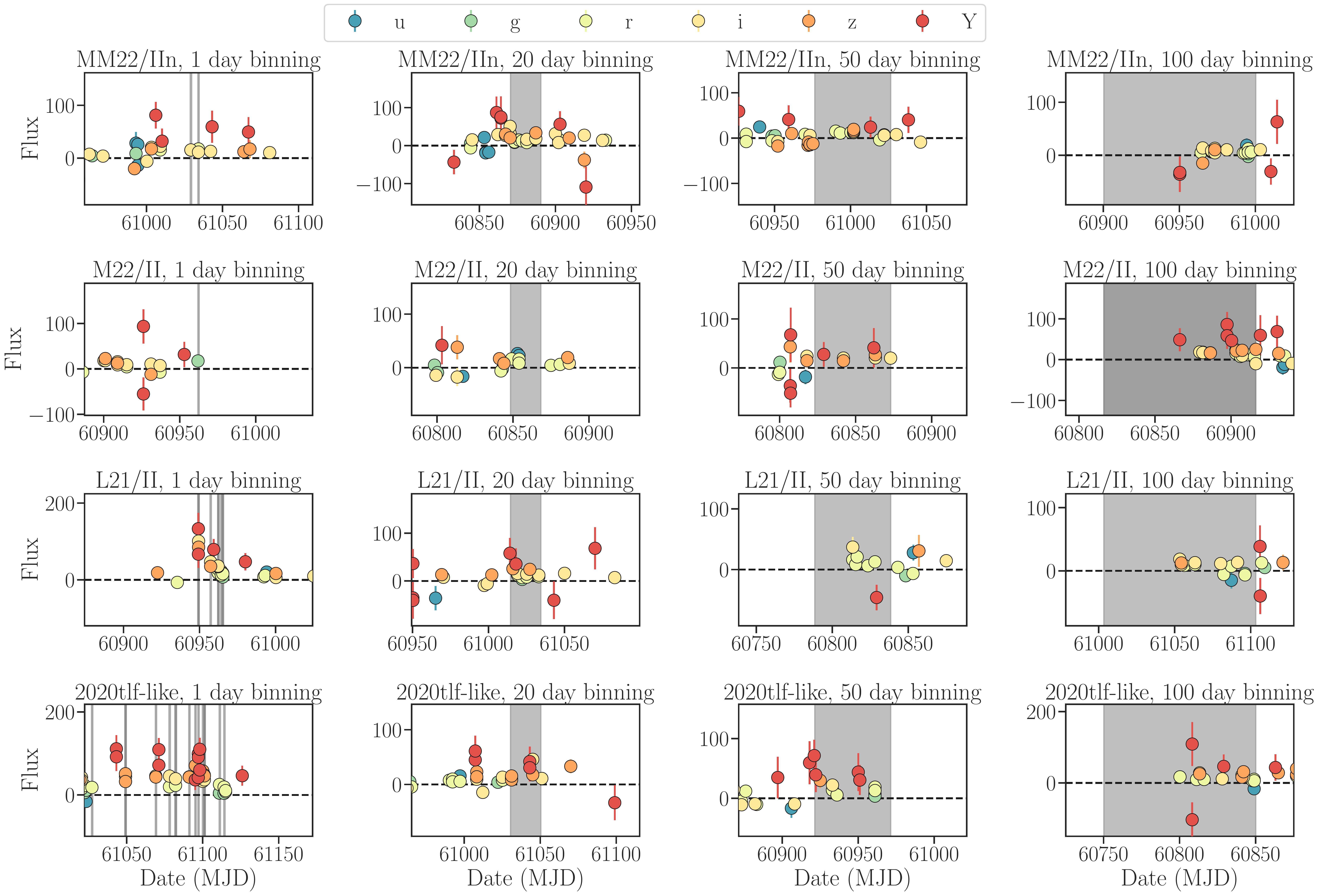}
\caption{Examples of binned precursor events with at least one 5-$\sigma$ detection for each model (rows) and bin width (columns), considering both Wide-Fast-Deep (WFD) and Deep Drilling Field (DDF) observations. Each bin in which a 5-$\sigma$ detection was made (in any band) is shaded. Most detections are made in LSST-$r$ and LSST-$i$.}
\label{fig:precursorExamples}
\end{figure*}

\end{document}